\documentclass[acmlarge,nonacm,screen]{acmart}
\AtBeginDocument{%
  }

\setcopyright{cc}
\setcctype{by}
\acmJournal{IMWUT}
\acmYear{2027}
\acmVolume{0}
\acmNumber{0}
\acmArticle{0}
\acmMonth{0}
\acmDOI{XXXXXXX.XXXXXXX}

\usepackage{acro}
\usepackage{array}
\DeclareAcronym{llm}{
    short=LLM,
    long=Large Language Model,
}
\DeclareAcronym{mvc}{
    short=MVC,
    long=Model-View-Controller,
}
\DeclareAcronym{bdar}{
    short=BDAR,
    long=Bidirectional Domain-Adaptive Representation,
}
\DeclareAcronym{cmda}{
    short=CMDA,
    long=Context-mediated domain adaptation,
}
\DeclareAcronym{cvc}{
    short=CVC,
    long=Context-View-Controller,
}
\DeclareAcronym{aco}{
    short=ACO,
    long=Adaptive Context Object,
}
\DeclareAcronym{ai}{
    short=AI,
    long=Artificial Intelligence,
}
\DeclareAcronym{ml}{
    short=ML,
    long=Machine Learning,
}
\DeclareAcronym{hci}{
    short=HCI,
    long=Human-Computer Interaction,
}
\DeclareAcronym{api}{
    short=API,
    long=Application Programming Interface,
}
\DeclareAcronym{iui}{
    short=IUI,
    long=Intelligent User Interfaces,
}
\DeclareAcronym{acm}{
    short=ACM,
    long=Association for Computing Machinery,
}

\AtBeginDocument{%
}

\newcommand{\varSouffleur}{\textsc{Souffleur}}
\newcommand{\condChat}{\textsc{Chat}}

\begin{document}

\title{Souffleur: A Glanceable Display for Mobile Voice Conversations}

\author{Anton Wolter}
\email{wol@cs.au.dk}
\orcid{0009-0004-6312-3355}
\affiliation{
  \institution{Aarhus University}
  \city{Aarhus}
  \country{Denmark}
}

\author{William Benjamin} 
\email{william.j.benjamin.phd@gmail.com}
\orcid{0009-0002-9548-3895}
\affiliation{
  \institution{Independent Researcher}
  \city{Chicago}
  \country{USA}
}

\author{Niklas Elmqvist}
\email{elm@cs.au.dk}
\orcid{0000-0001-5805-5301}
\affiliation{
  \institution{Aarhus University}
  \city{Aarhus}
  \country{Denmark}
}

\renewcommand{\shortauthors}{Wolter et al.}

\begin{abstract}
    The speech capabilities of mobile devices have grown, but their displays still rely on a text-based chat.
When resuming a voice chat after a context switch, the last messages may not convey the key points of the conversation.
Autobiographical use and sessions with external participants led to design goals for a mobile-first, speech-driven, glanceable display that helps users continue their reasoning from where they left off.
We implement these goals in ``Souffleur'' and evaluate it in a between-subjects study with 48 participants.
Compared with a conventional chat transcript, we found no reliable difference in how easily participants followed and resumed short voice conversations, nor in effort, attention or recall, although Souffleur reduces the conversation to a few short phrases.

\end{abstract}

\begin{CCSXML}
<ccs2012>
   <concept>
       <concept_id>10003120.10003138.10003140</concept_id>
       <concept_desc>Human-centered computing~Ubiquitous and mobile computing systems and tools</concept_desc>
       <concept_significance>500</concept_significance>
       </concept>
   <concept>
       <concept_id>10003120.10003145.10003147.10010923</concept_id>
       <concept_desc>Human-centered computing~Information visualization</concept_desc>
       <concept_significance>300</concept_significance>
       </concept>
   <concept>
       <concept_id>10003120.10003121.10003124.10010865</concept_id>
       <concept_desc>Human-centered computing~Graphical user interfaces</concept_desc>
       <concept_significance>100</concept_significance>
       </concept>
   <concept>
       <concept_id>10003120.10003121.10003124.10010870</concept_id>
       <concept_desc>Human-centered computing~Natural language interfaces</concept_desc>
       <concept_significance>300</concept_significance>
       </concept>
 </ccs2012>
\end{CCSXML}

\ccsdesc[500]{Human-centered computing~Ubiquitous and mobile computing systems and tools}
\ccsdesc[300]{Human-centered computing~Information visualization}
\ccsdesc[100]{Human-centered computing~Graphical user interfaces}
\ccsdesc[300]{Human-centered computing~Natural language interfaces}

\keywords{
  voice interaction,
  conversational agents,
  glanceable displays,
  mobile computing,
  interruption and resumption,
  information visualization
 }

\begin{teaserfigure}
    \centering
    \includegraphics[width=0.6\textwidth]{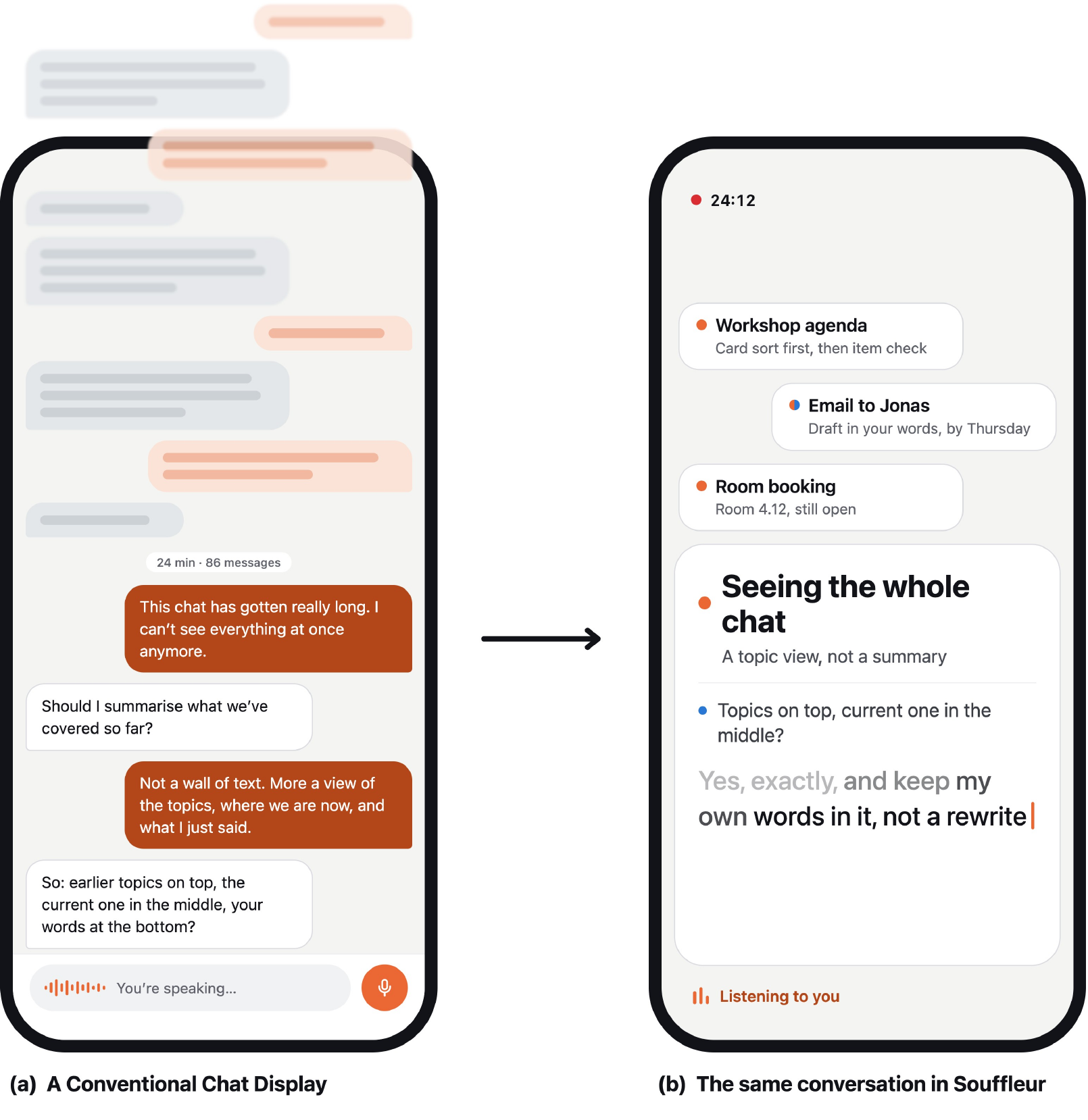}
    \caption{
      \textbf{Souffleur.}
      (a) a conventional chat transcript with an overflowing chat history, indicating that the full information cannot be displayed within the available screen size.
      The further the messages are out of the viewport, their opacity reduces and blur increases, indicating that the information may become blurry as well.
      (b) the same conversation as a topic stack, with the current topic in the large card, earlier topics above it, each with a decision line or open status, orange and blue for the two speakers, and a status line at the bottom.
    }
    \Description{
      Two phone screens side by side.
      Left: a conventional chat transcript with alternating messages.
      Right: the Souffleur topic stack, with a large current-topic card titled ``Seeing the whole chat'' and the line ``A topic view, not a summary'', above earlier topic bubbles for ``Workshop agenda'', ``Email to Jonas'' and ``Room booking'', with the status line ``Listening to you''.
    }
    \label{fig:teaser}
\end{teaserfigure}

\maketitle

\section{Introduction}

Thinking aloud is one of the oldest ways to work through a problem: 
explaining it to a colleague, talking it through on a walk, talking a half-formed idea into shape.
Voice agents built on \acp{llm} now offer a capable partner for this whenever the hands and eyes are busy with something else, such as walking, cooking or commuting.
In these moments, the conversation is a secondary task and the attention belongs to the real world, while the phone is at most something to glance at.

Nevertheless, the display that comes with these conversations is still the chat transcript.
Assistant voice modes show every turn as a message, the interface people know from texting.
A transcript works while a conversation is short, but after a few minutes of speech it becomes a column of text that cannot be taken in at a glance.
Furthermore, spoken language reads poorly on the page, with its fragments, restarts and fillers.
Speech itself is linear and transient~\cite{kim2011usabilityproblemsinvoicesystems,leahy2011clt_transient}.
Earlier ideas have to be held in working memory, the resource the speaker also spends on producing the next idea and on the task at hand~\cite{wickens2008multipleresources}.

The problem is sharpest when the conversation is interrupted, which in mobile settings is the rule rather than the exception: a light turns green, a call comes in or the mind just wanders off by itself.
Resuming a suspended task depends on how quickly its goal comes back to mind, and cues in the environment speed this up~\cite{altmann2002memoryforgoals,trafton2003preparingtoresume}.
In text chat, the scrollback provides such cues.
In a voice conversation, the speaker has at best the last thing the agent said, or a transcript they would have to read.
Our own formative use made this concrete.
After a context switch, the last messages rarely conveyed where the conversation stood, what had been decided and what was still open.
People remember what they generated or said themselves better than what they were given or heard~\cite{slamecka1978thegenerationeffect,forrin2018hearingoneself}, and they lose track of which ideas came from an AI~\cite{zindulka2026aimemorygap}.

Existing systems address neighbouring problems.
Writing tools turn speech into persistent text~\cite{lin2024rambler,elalaoui2025stepwrite}, and memory aids retrieve what was said in the past~\cite{zulfikar2024memoro}.
Neither shows where the current conversation stands.
Speech-to-canvas systems such as Orality~\cite{li2026orality} and VoCa~\cite{ge2026vocadesigningspeechcanvasinteraction} build rich visual structure beside a spoken conversation, made to be read and explored, and VoCa's participants found it hard to follow the speech and the canvas at once~\cite{ge2026vocadesigningspeechcanvasinteraction}.
Meeting tools visualise the topics of group conversations on shared screens~\cite{chandrasegaran2019talktraces,chen2025meetmap}.
Studies of screens for voice assistants show that a screen can lower workload~\cite{chen2023screenornoscreen} and that minimal visual feedback is preferred when the eyes are busy~\cite{baghdadi2025richhud}, but these studies concern single commands, not an open conversation with an \ac{llm}.
We lack an account of which minimal representation of an ongoing voice conversation helps a person at a glance on a phone: to follow it, to pick it up again after an interruption, and to come away knowing what was decided.

We address this gap with \textit{Souffleur}, a voice-first thinking partner with a glanceable, mobile-first display.
Souffleur is named after the theatre prompter who stays silent while the actor speaks and gives a cue only when one is needed.
The conversation stays in speech. 
The screen is a companion to glance at, not a second place to read.
Souffleur folds the conversation into short topic headlines and marks two things on each: whose words the topic mostly rests on, the person's or the agent's, and whether it is settled or still open.
We explored several ways to lay out these topics. 
This paper focuses on the \textit{topic stack}, which keeps each topic in a fixed slot.

The design follows from formative work: four months of autobiographical use and research through design, and sessions with external participants.
From these we derive design goals for glanceable displays of voice conversations that are held as a secondary task (\autoref{sec:formative}).
We then ask:

\begin{itemize}
    \item \textbf{RQ1:} How does a glanceable topic display of an ongoing voice conversation compare with a transcript in how easily people follow the conversation, and what does each cost them in effort and attention?
    \item \textbf{RQ2:} How does the topic display compare with a transcript in how fast and accurately people resume a voice conversation after an interruption?
    \item \textbf{RQ3:} How does the topic display affect what people take away from the conversation, and what on the screen helps or gets in the way?
\end{itemize}

We answer these questions with a between-subjects study.
In it, 48 participants each held one short voice conversation with an \ac{llm} agent about an everyday planning or reflection task, on their phone with either the topic stack or a plain transcript as a control.
Partway through each conversation, a short game took over the screen and the agent paused, and we measured how people resumed: the time to their first words afterwards, and whether those words picked up where the conversation had been.
After each conversation, participants summarised it from memory, and a blind \ac{llm} judge scored the summaries against the conversation.
We found no reliable difference between the two displays.
With both, participants rated the conversation as easy to follow, resumed almost always on the topic they had left, and remembered a similar share of key points.
The study could detect only large differences, and participants held the conversation without a concurrent task, so these results bound rather than settle the value of a glanceable display.

This paper makes the following contributions:
\begin{enumerate}
    \item Formative insights and six design goals for glanceable displays that accompany voice conversations held as a secondary task.
    \item Souffleur, a voice-first thinking partner whose display folds the conversation into a glanceable topic stack that marks provenance and status.
    \item A between-subjects study with 48 participants comparing the topic stack with a transcript on following, resuming and recalling the conversation.
    \item Design implications for displays of voice conversations that people glance at rather than read.
\end{enumerate}
 
\section{Related Work}
\label{sec:related}

Our work sits where four lines of research meet: systems that let people think aloud with a voice agent, voice interaction as a secondary task, interruption and resumption, and the visualisation of conversation.

\subsection{Thinking Aloud with Voice Agents}
\label{sec:rw-thinking}

Capturing spoken thought is an old goal.
VoiceNotes~\cite{stifelman1993voicenotes} let people record and organise spoken notes on a hand-held device long before smartphones.
Speaking a thought forces it into a new form, which makes it an epistemic action~\cite{kirsh1994epistemicpragmaticaction} rather than only a report, and saying something aloud improves memory for it~\cite{macleod2010theproductioneffect}.
\acp{llm} add a partner that can answer, question and restructure what is said.
Rambler~\cite{lin2024rambler} turns dictated speech into gists that writers manipulate on screen, StepWrite~\cite{elalaoui2025stepwrite} scaffolds voice-first writing, TalkSketch~\cite{shi2026talksketch} combines speech with design sketching, and GazeNoter~\cite{tsai2025gazenoter} uses gaze to refine notes during a conversation.
Memoro~\cite{zulfikar2024memoro} retrieves the user's earlier memories through a deliberately concise interface, Talk2Care~\cite{yang2024talk2care} structures what older adults tell a voice assistant for their healthcare providers, and Memolet~\cite{yen2024memolet} gives users a visual handle on earlier AI conversations in typed chat.

Two recent systems come closest to ours.
Orality~\cite{li2026orality} turns speech into a semantic node-link canvas for externalising and clarifying thoughts.
VoCa~\cite{ge2026vocadesigningspeechcanvasinteraction} pairs a voice \ac{llm} agent with a canvas of notes, diagrams and discussion frames.
It was deployed for five days with 18 participants, who found it hard to follow the speech and the canvas at once, and the canvas lagged behind what was said.
Both systems treat the display as a rich second workspace that rewards close reading.
Our formative work led us to the opposite stance: 
the conversation stays in speech, and the display has to work in the time of a glance.

\subsection{Voice Interaction as a Secondary Task}
\label{sec:rw-secondary}

Talking is not free while doing something else.
Holding a phone conversation impairs simulated driving even hands-free~\cite{strayer2001distraction}, and multiple-resource theory predicts the strongest interference between tasks that draw on the same resources, such as two verbal ones~\cite{wickens2008multipleresources}.
An on-road study found that a multi-turn conversation with an \ac{llm}-based in-car agent imposed a load comparable to a hands-free phone call~\cite{monk2026incarllm}.
Speech-only interfaces add problems of linearity, transience, ambiguity and minimal feedback~\cite{kim2011usabilityproblemsinvoicesystems}, and long spoken explanations overload working memory because they cannot be revisited~\cite{leahy2011clt_transient}.
Turn-taking adds a further hazard for thinking aloud: systems that end a turn at a silence threshold cannot tell a thinking pause from the end of a turn~\cite{skantze2021turntaking}, and current full-duplex speech models take the floor during many user pauses~\cite{lin2025fullduplexbench}.

A screen can relieve some of this load.
In a deployment over 40 days with older adults, a voice assistant with a touchscreen lowered workload compared with the same assistant without one, while most answers were still given by voice~\cite{chen2023screenornoscreen}.
In a driving simulator, a head-up display showing only key words was rated more usable than two richer displays for an in-vehicle voice assistant~\cite{baghdadi2025richhud}.
Hey Dashboard!~\cite{dhanoa2026heydashboard} uses \acp{llm} to support voice, text and pointing in dashboard onboarding.
Visual backchannels can show that a voice interface is listening~\cite{park2025visualbackchannels}, and brief on-screen text lets users check what an agent heard and repair it~\cite{moore2024understandingtwowaystreet}.
Other work focuses on displaying text during multitasking in mobile settings, such as GlassMessaging~\cite{janaka2023glassmessaging}, a messaging interface for smart glasses, where participants reply faster but under higher workload than in a phone chat.
LSVP~\cite{ram2021onthegovideolearning} showed that presenting content sequentially and keeping earlier elements on screen helped people learn from video while walking.
We aim to show the state of an open, multi-topic conversation, with a focus on glanceability after a person's attention has been elsewhere.

\subsection{Interruption and Resumption}
\label{sec:rw-interruption}

Mobile and everyday settings lead to fragmented attention.
Information workers switch between activities every few minutes~\cite{gonzalez2004multitasking}.
Interrupted work is completed faster, but at the cost of more stress and effort~\cite{mark2008interruptedwork}.
After an alert, people often take many minutes to return to the task they suspended~\cite{iqbal2007disruption}.
The memory-for-goals account explains why resuming is hard: 
a suspended goal decays while attention is elsewhere and has to be brought back to mind, and cues in the environment speed up its retrieval~\cite{altmann2002memoryforgoals}.
Being warned before an interruption and rehearsing the goal both shorten the time it takes to resume~\cite{trafton2003preparingtoresume}.
Proactive voice agents have to choose opportune moments to speak, and whether a moment is opportune depends on how busy and engaged the person is~\cite{cha2020opportunemoments}.
In an on-road study, drivers' interruptibility for a voice task depended on the task's demand and on the drive.
Most adapted by slowing down or by adjusting their engagement with the voice task~\cite{kim2020interruptibilityinvehiclemultitasking}.

For a voice conversation, these findings suggest that the display's main job after an interruption is to act as a resumption cue.
A full transcript offers cues, but only to someone who reads all of it, while the last utterances offer too little.
We are not aware of work that measures how people resume a conversation with a voice agent, or how a display affects it.

\subsection{Visualising Conversation}
\label{sec:rw-visualising}

Social visualisations show who is present and how much each person talks: Chat Circles~\cite{viegas1999chatcircles} shows participants as circles that grow with each new message, and the Conversation Clock~\cite{bergstrom2007conversationclock} shows the audio contributions of a co-located group on a table.
These displays are glanceable, but they show participation, not content.
Content visualisations show what is said, such as TalkTraces~\cite{chandrasegaran2019talktraces}, which captures a meeting's speech in real time and visualises its topics, and MeetMap~\cite{chen2025meetmap} uses \acp{llm} to build dialogue maps for online meetings.
For typed \ac{llm} chat, Sensecape~\cite{suh2023sensecape} and Graphologue~\cite{jiang2023graphologue} restructure responses into hierarchies and diagrams, and OnGoal~\cite{coscia2025ongoal} tracks which of a user's goals a multi-turn conversation has addressed.
These content views are made to be read, usually on large screens, and their evaluations concern usefulness and perceived load rather than how people resume.

Glanceability is well defined for peripheral displays:
a glanceable display lets people take in information quickly and with little effort, at the edge of their attention~\cite{matthews2006glanceable}.
Our display applies this to the content of a conversation.
It shows a few topics and two signals per topic, for one person talking with an AI on a phone, and we evaluate it on following, resuming and recalling the conversation.

\section{Formative Work and Design Goals}
\label{sec:formative}

Souffleur's display grew out of four months of building and using a voice thinking partner in everyday settings. 
This section derives six design goals from what we observed, each grounded in the literature that explains why the observation generalises.

\subsection{Formative Activities}
\label{sec:formative-activities}

We followed a research-through-design process~\cite{zimmermann2007researchthroughdesign}, reporting mostly the authors' own use as autobiographical design~\cite{neustaedter2012autobiographicaldesign}, complemented by a handful of external participants over multiple weeks.
The external participant we cite most was a relative of the first author, a native German speaker who used the system in twelve sessions over nine days.
We translated her quotes from German.
Because of this relationship, we treat her sessions as formative only.

The system evolved across three conceptual framings: from an AI sketching canvas, to an eyes-free cognitive prompter (\textit{Souffleur}), and to a voice-reifiable command environment.
Each prototype served as a design probe for the tensions between generative voice models and human cognition during eyes-busy, hands-busy multitasking.
This paper reports the display that grew out of the second framing. The third separated commands from content, after early headless prototypes suffered a vocal ``Midas Touch'' problem in which spoken commands were mistaken for content, and is not evaluated here.

\subsubsection{Learning 1: Externalisation Over Visual Synthesis}
Our first prototype coupled live speech with sequential canvas sketching (\texttt{tldraw}), hoping that visual mind maps would support spoken ideation. 
Early testing revealed two limitations. 
First, turning speech into visuals largely repackaged what foundation models already did. 
Second, inspecting a canvas or a noisy transcript took attention that mobile, eyes-free use did not leave.

Setting the canvas aside, we held over forty voice-only sessions while developing a major research grant proposal, and characterised spoken ideation as ``building a ship inside a bottle''.
Speakers must externalise complex, interconnected mental models through the narrow, sequential channel of speech, in fragments and revisions across sessions. 
Once sufficient conceptual structure was externalised and preserved, generating polished written artefacts became comparatively straightforward.
The bottleneck was conceptual externalisation, not content generation.

This led to a minimally intrusive, hands-free thinking partner that preserves emerging ideas and steps in without disrupting the speaker's train of thought.

\subsubsection{Learning 2: Conversational Discipline and the \textit{Souffleur} Metaphor}
To establish baselines, we evaluated commercial voice assistants and passive recorders during driving commutes.
Long spoken answers overloaded working memory, while passive voice recorders offered no feedback, creating a ``brick wall'' that pressured us to keep filling the silence.
The challenge was to provide enough support without interrupting the flow of thought.

We adopted the theatrical metaphor of a \textit{souffleur}, a prompter who remains silent while an actor performs, offering brief cues only when needed.
We turned this into a strict conversational discipline:
answers to direct questions were one sentence of under 25 words, and the agent defaulted to an explicit \texttt{<silence>} token while the person elaborated.
Silence thus became a deliberate design mechanism, preserving cognitive continuity and user agency.

\subsubsection{Learning 3: Glanceable Semantic Anchors and Cross-Session Resumption}

Using voice on the move revealed a challenge:
interruptions often broke the train of thought, and fleeting speech offered few cues for resuming.
Rather than a demanding transcript, we showed a real-time semantic header: 
a short, glanceable topic phrase as an anchor whenever the eyes were free.

An unexpected state-persistence bug revealed a second opportunity.
When the header stayed visible across sessions, seeing the previous topic appeared to revive unfinished ideas and encourage their continuation.
We recognised the bug as an opportunity and kept the semantic anchor across sessions.
This turned the interface from a session-bound voice recorder into a persistent cognitive workspace~\cite{klokmose2015webstrates}, supporting both recovery from interruptions and the incubation of ideas over time.

\subsection{Design Goals}
\label{sec:design-goals}

\begin{figure*}[t]
    \centering
    \begin{minipage}[t]{0.32\textwidth}
        \centering
        \includegraphics[width=\linewidth]{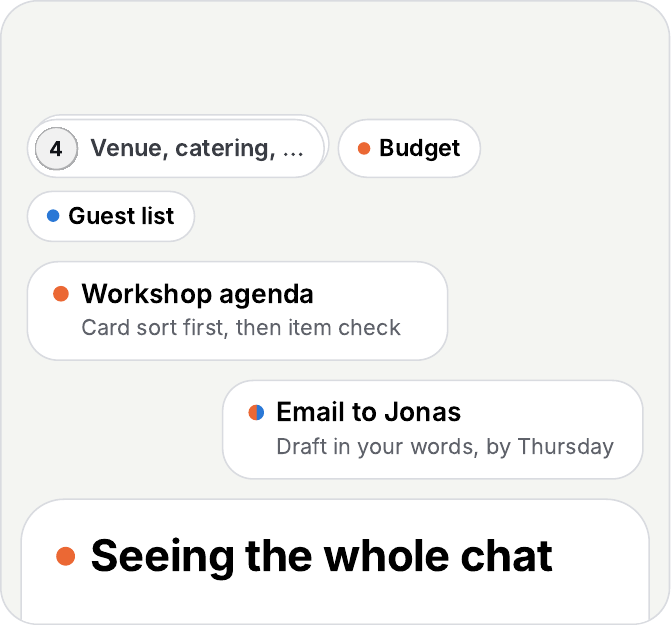}\\[2pt]
        \par\raggedright\small\textbf{(a) DG1 Glance, don't read.} 
        Few short elements.
        Older topics fold into name-only pills and clusters.
    \end{minipage}\hfill
    \begin{minipage}[t]{0.32\textwidth}
        \centering
        \includegraphics[width=\linewidth]{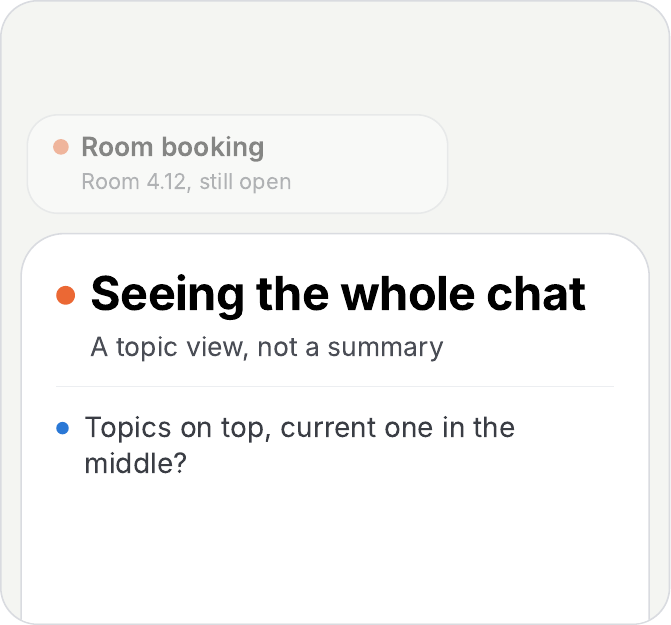}\\[2pt]
        \par\raggedright\small\textbf{(b) DG2 Show the suspended goal.} 
        The current topic is the largest element, with what has been decided on it.
    \end{minipage}\hfill
    \begin{minipage}[t]{0.32\textwidth}
        \centering
        \includegraphics[width=\linewidth]{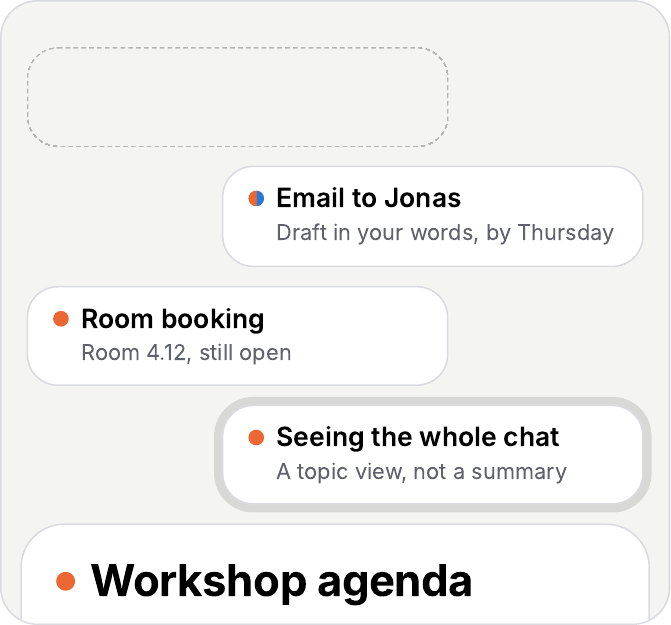}\\[2pt]
        \par\raggedright\small\textbf{(c) DG3 Keep everything in its place.} 
        Each topic keeps its slot.
        A topic that comes back leaves a gap.
    \end{minipage}\\[10pt]
    \begin{minipage}[t]{0.32\textwidth}
        \centering
        \includegraphics[width=\linewidth]{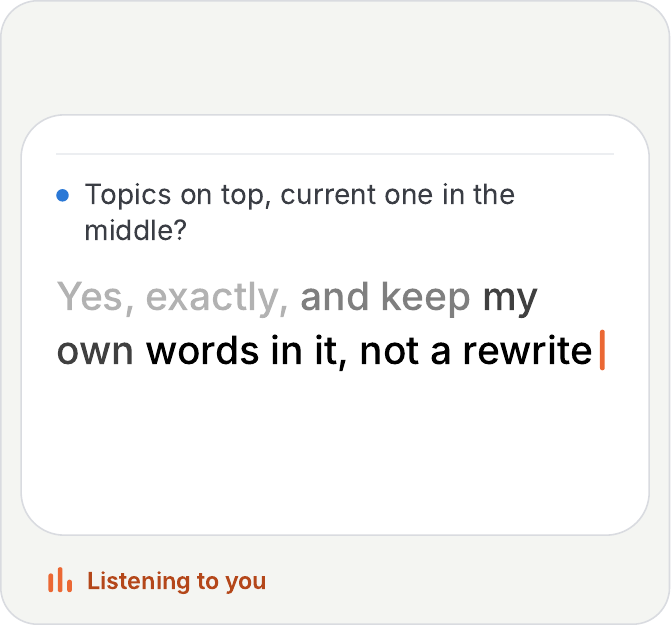}\\[2pt]
        \par\raggedright\small\textbf{(d) DG4 Overview first, the current words in view.} 
        The turn being answered as a gist, the words being said as they come.
    \end{minipage}\hfill
    \begin{minipage}[t]{0.32\textwidth}
        \centering
        \includegraphics[width=\linewidth]{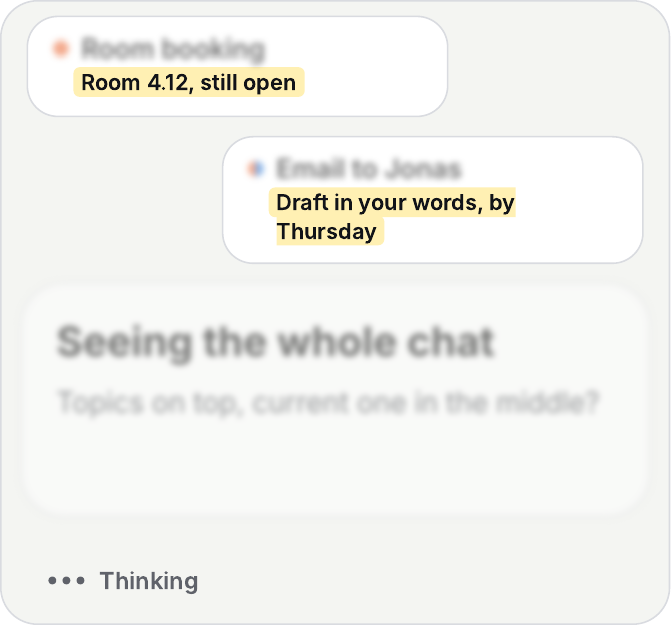}\\[2pt]
        \par\raggedright\small\textbf{(e) DG5 Show the state, not only the content.} 
        A decision line on each topic, and one line under the conversation for what the system is doing.
    \end{minipage}\hfill
    \begin{minipage}[t]{0.32\textwidth}
        \centering
        \includegraphics[width=\linewidth]{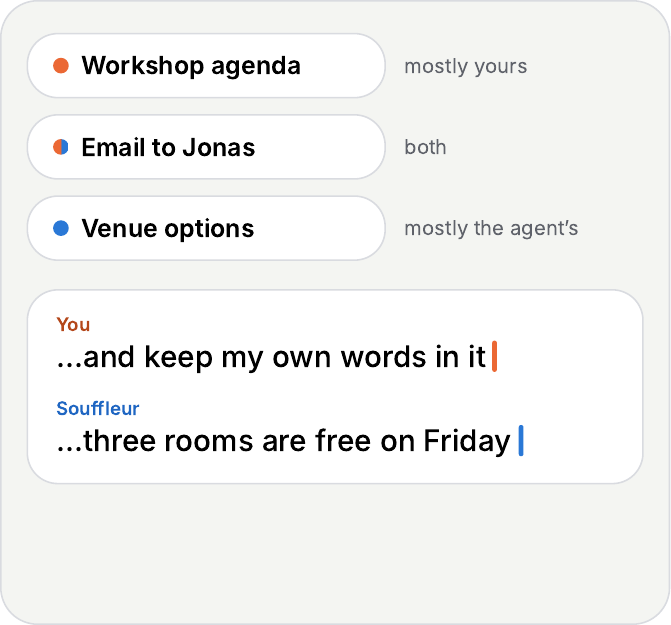}\\[2pt]
        \par\raggedright\small\textbf{(f) DG6 Make the AI's part visible.} 
        Orange for the person, blue for the agent, from the live words up to the topics.
    \end{minipage}
    \caption{Souffleur's topic stack, cropped to the feature that meets each design goal. All crops come from the same conversation as \autoref{fig:teaser}, at different moments. In (e), the topic names are blurred to bring out the decision lines.
    In (f), labels name the three provenance states.}
    \Description{Six cropped phone screens in a three-by-two grid. (a) Earlier topics as two bubbles with a decision line each, above three small pills, one of them a cluster with the count four. (b) A large white card titled ``Seeing the whole chat'' with the line ``A topic view, not a summary''. (c) Earlier topics on their slots: a dashed empty slot where a topic came back from, and the topic just left with a soft grey ring, above the current topic ``Workshop agenda''. (d) A gist line marked with a blue dot above large live words whose first words are faded, ending in an orange caret, and the line ``Listening to you''. (e) Two topic bubbles whose names are blurred and whose decision lines are highlighted, above a blurred current topic and the status line ``Thinking''. (f) Topic bubbles with an orange, a half orange and half blue, and a blue dot, and two lines of live words ending in an orange and a blue caret.}
    \label{fig:design-goals}
\end{figure*}

\autoref{fig:design-goals} shows how Souffleur's display meets each goal.

\begin{description}
    \item[DG1: Glance, don't read.]
    \textit{Observed.}
    Our first prototype drew a canvas beside the conversation, and inspecting it, like reading a noisy transcript, took attention that walking or driving did not leave.
    Long spoken answers from commercial assistants overloaded working memory.
    The more the app showed, the less it helped:
    on an evaluation drive it was ``too many different screens'' and ``all of it is too much for me'', and an external participant asked for ``everything on one screen''.
    A bar under each topic was read as ``how much do I understand'', because its meaning was hidden.
    In our first pilots of the topic stack, the live words grew until no earlier topic was visible on a small phone.
    \textit{Grounded.}
    A glanceable display conveys its information quickly and with little effort, at the edge of attention~\cite{matthews2006glanceable,srinivasan2025attentionaware}.
    While a person speaks, the verbal channel is already taken and reading competes with it~\cite{wickens2008multipleresources}, and with the eyes busy, a display of key words is preferred over richer ones~\cite{baghdadi2025richhud}.
    VoCa's participants found it hard to follow speech and a canvas at once~\cite{ge2026vocadesigningspeechcanvasinteraction}.
    \textit{Therefore} the display shows few elements in short phrases, and every mark explains itself.
    Each part has a fixed budget, so none can crowd out another as the conversation grows: older topics fold into name-only pills and then into clusters (\autoref{fig:design-goals}a).

    \item[DG2: Show the suspended goal.]
    \textit{Observed.}
    A header naming the current topic in a few words became the anchor for getting back into the conversation after an interruption. 
    When it stayed visible into the next session, it appeared to revive unfinished ideas.
    The advisor asked for ``a short recap on re-entry into a topic'', and on an evaluation drive the agent was asked ``How can I see my last discussion?'', which the conversation screen could not answer.
    \textit{Grounded.}
    Resuming an interrupted task means bringing its goal back to mind, and cues in the environment speed this up~\cite{altmann2002memoryforgoals,larsenledet2025traces,trafton2003preparingtoresume}.
    The longer and more demanding the interruption, the slower people resume~\cite{monk2008interruptionduration}.
    \textit{Therefore} the topic being talked about, and what has been decided on it, is the largest element of the display (\autoref{fig:design-goals}b).

    \item[DG3: Keep everything in its place.]
    \textit{Observed.}
    A display with one panel per speaker scattered the order of the conversation: ``some of the things that I said on the bottom right \ldots{} it's going left, right''.
    A return to an earlier topic put the latest lines under its old heading, ``on top of the transcript''.
    \textit{Grounded.}
    Interfaces that rearrange themselves are slower than stable ones~\cite{findlater2004menus}, and adaptations pay off only when people can predict them~\cite{gajos2008predictability}.
    In the periphery of vision, motion is noticed where changes of colour are missed~\cite{bartram2003moticons}, and a change that happens while the eyes are elsewhere often goes unseen~\cite{simons2005changeblindness}.
    Animated transitions help people follow a change when they are staged and short~\cite{chalbi2020commonfate,heer2007animatedtransitions}.
    \textit{Therefore} the layout is fixed and each topic keeps its slot once it has one.
    A topic that comes back leaves a gap, so no other topic changes place (\autoref{fig:design-goals}c).
    Only the oldest topics change, when they fold into pills (DG1).
    The return itself is the only animated move, and it is marked afterwards for a glance that missed it (\autoref{fig:return}).

    \item[DG4: Overview first, the current words in view.]
    \textit{Observed.}
    A raw transcript of speech was close to useless on the page, yet the words were needed to check what the system had heard:
    the recogniser never produced an external participant's Low German song title, and only her spoken corrections brought the right title into her list.
    Words that appeared only after a turn had ended left the speaker with no sign of being heard while speaking.
    \textit{Grounded.}
    Information displays should show an overview first and details on demand~\cite{shneiderman1996eyeshaveit}, and overview and detail can be shown together or a step apart~\cite{cockburn2009overviewdetail}.
    Brief on-screen text lets people check what a conversational agent heard and repair it~\cite{moore2024understandingtwowaystreet}.
    Captions that revise themselves are uncomfortable to follow~\cite{liu2023captionstability}, and keeping earlier elements on screen helps people take in content while on the move~\cite{ram2021onthegovideolearning}.
    \textit{Therefore} the topics are the overview, and only the current turn appears as words: the last few words being said, appended rather than revised word by word, under a gist of the turn being answered (\autoref{fig:design-goals}d).
    Nothing is lost: a tap on a topic opens what each side said on it, and the words themselves one tap further.

    \item[DG5: Show the state, not only the content.]
    \textit{Observed.}
    One external participant could not find the song list she had built over four sessions:
    she asked where something was seven times in two sessions, while the list existed in seven places.
    The advisor wanted the system to propose what to talk about, fitted to the time available.
    Silence from a passive recorder created a ``brick wall'' (\autoref{sec:formative-activities}): without feedback, we could not tell whether the system was still listening.
    \textit{Grounded.}
    Awareness of a situation goes beyond seeing its parts to understanding what they mean together~\cite{endsley1995situationawareness}, and visual backchannels show that a voice interface is listening~\cite{park2025visualbackchannels}.
    \textit{Therefore} each topic carries a line of what was decided on it, or what is still open, and one line under the conversation says what the system is doing: listening, thinking, looking something up, speaking or waiting (\autoref{fig:design-goals}e).

    \item[DG6: Make the AI's part visible.]
    \textit{Observed.}
    The structure of an external participant's workspace was entirely the system's: of 75 workspace operations, 68 came from automatic extraction and 7 from the agent, none from her own hand, and she edited none of the six drafts the agent wrote of her song list.
    \textit{Grounded.}
    People lose track of which ideas came from an AI, most of all ideas they elaborated themselves~\cite{zindulka2026aimemorygap}, do not feel they own AI-generated text~\cite{draxler2024aighostwritereffect}, and claim ideas they improved on as their own~\cite{stark2005appropriation}.
    \textit{Therefore} one colour marks the person's words and another the agent's, wherever words appear: on the caret of the live words, on the gist of a turn, and on each topic, which shows whether it rests mostly on the person, mostly on the agent, or on both (\autoref{fig:design-goals}f).
\end{description}

\section{Implementation: Souffleur}
\label{sec:implementation}

Souffleur runs as a progressive web app on the person's own phone.
It has three parts: the voice conversation with the agent, a pipeline that folds the conversation into topics as it goes, and the display that shows them, the \emph{topic stack}.
This section describes each and how the study is instrumented.
The design goals (\autoref{sec:design-goals}) set the budgets and timings given here.

\subsection{System Architecture}

Souffleur is a TypeScript monorepo.
A Next.js\footnote{\href{https://github.com/vercel/next.js/}{https://github.com/vercel/next.js/} last accessed Oct 27 2026} web application serves the app and its API routes.
A worker processes jobs from a queue in the same PostgreSQL\footnote{\href{https://www.postgresql.org/}{https://www.postgresql.org/} last accessed Oct 30 2026} database: it transcribes audio, runs extraction and exports data.
All language-model calls apart from the voice conversation itself go through a LiteLLM proxy\footnote{\href{https://github.com/BerriAI/litellm}{https://github.com/BerriAI/litellm} last accessed Oct 31 2026}, which lets us choose a model per role (transcription, titling, fast summaries, judge) without changing code.
The system is deployed with Docker on a self-hosted Coolify\footnote{\href{https://github.com/coollabsio/coolify}{https://github.com/coollabsio/coolify} last accessed Oct 30 2026} instance.

Every word is kept in an append-only ledger.
The phone records the person's audio in chunks of 10\,s, queues them locally and uploads them.
Then the worker transcribes each chunk with a self-hosted Whisper model and deletes the audio once it is transcribed.
The agent's turns are stored separately, with their timing.
For analysis, PostHog\footnote{\href{https://github.com/PostHog/posthog}{https://github.com/PostHog/posthog} last accessed Oct 30 2026} logs interaction events with session replays in which all text is masked, and Langfuse\footnote{\href{https://github.com/langfuse/langfuse}{https://github.com/langfuse/langfuse} last accessed Oct 29 2026} keeps one trace per agent turn.
Study events carry only counts, timings and identifiers, never what was said.

\subsection{The Voice Conversation}

The conversation runs on OpenAI's \texttt{gpt-realtime-2.1} speech-to-speech model\footnote{\href{https://platform.openai.com/docs/guides/realtime}{https://platform.openai.com/docs/guides/realtime} last accessed Oct 31 2026}, connected from the browser over WebRTC.
The server only composes the agent's instructions and mints a key for one session, so audio does not pass through our servers on its way to the model.
The model decides when a turn has ended by semantic voice activity detection, which waits longer when the person sounds mid-thought than a silence threshold would~\cite{skantze2021turntaking}.
On our test drives, the time from the end of a turn to the agent's first sound was 0.7 to 1.3\,s.
An earlier pipeline built with Pipecat\footnote{\href{https://github.com/pipecat-ai/pipecat}{https://github.com/pipecat-ai/pipecat} last accessed Oct 31 2026}, with separate speech recognition, language model and speech synthesis, is kept as a fallback.
It ended a third of the turns in a pilot only by a 3\,s timeout, which is why we moved to the speech-to-speech model.

The agent's instructions adapt the conversational discipline of our formative work (\autoref{sec:formative-activities}).
Where the formative agent answered in one sentence of under 25 words and stayed quiet by default, the study agent answers in at most 60 words, usually two or three spoken sentences, brings its own ideas and hands back one short question.
It never summarises the conversation or lists what has been covered, and it does not mention the screen: the structure of the conversation lives on the display, not in the agent's voice.
While the person is audibly mid-thought, the agent may only say ``Mm.'' or ``Go on.''.

Two streams of text reach the display.
The person's words come from a second, transcription-only session on the same microphone track, which streams words while they speak and is committed when the conversation's turn ends.
The turn's final transcript then replaces the live words in place.
The agent's reply arrives as text before its audio, and WebRTC gives no timing per word.
Souffleur therefore lays out the reply's words in advance, invisibly, and reveals them at the pace of the voice, estimated at 15 characters per second. 
Each word fades in where it will stay, over 320\,ms.
A reply the person interrupts is cut to the part that was heard.

\subsection{Folding the Conversation into Topics}

Three language-model calls, each with its own update rule, turn the conversation into the content of the display.

\paragraph{Topics}
A titler names the current topic after every user turn that adds at least 20 characters, and every 30\,s while there is new speech.
It sees the last 900 characters of the conversation, the current title and up to 16 earlier titles, and returns the current title unchanged if the subject is the same, an earlier title if the talk has gone back to it, or a new title of two to four words for the most specific subject being discussed (\textit{DG1}).
A title that matches an earlier one counts as a return to that topic.
Each topic records how much of it each side said: a topic rests on the person if they said at least 65\% of its words, on the agent if it did, and on both otherwise (\textit{DG6}).

\paragraph{Decisions}
After 2.5\,s of silence, a digest call reads the last 80 lines and writes, for each topic, a decision line of at most six words with concrete times, places and numbers, such as ``Room 4.12, still open'' (\textit{DG5}).
It records only what the person agreed to, never a suggestion they did not take up, and names what is still unclear as unclear.
Topics are referred to by number in the call and matched back by number, so the model cannot add topics of its own.

\paragraph{Gists}
When a new turn begins, the turn before it is shrunk to a gist of at most eight words (\textit{DG4}).
Until the gist arrives, the display shows the turn's last question, or else its last sentence without fillers, drawn lighter.
Whichever gist arrives first, from this call or from the digest, is kept and never rewritten.

Titles use the fallback pipeline's language model at temperature~0, as the titling role has no model of its own.
Decisions and gists use a smaller, fast model, \texttt{gemma3:12b}, at temperature~0.2. 
Both are configured through the proxy.
None of these calls blocks the conversation: the agent speaks while the display catches up.

\subsection{The Topic Stack}

The display stacks the conversation from the oldest to the newest, top to bottom (\autoref{fig:teaser}, \autoref{fig:design-goals}).
At the bottom, one status line says what the system is doing: \textit{Listening to you}, \textit{Thinking}, \textit{Looking it up}, \textit{Speaking} or \textit{Waiting for you}, with \textit{Connecting} while the session opens and a \textit{Locked} note when the agent talks only when poked (\textit{DG5}).
Above it, the current topic is a large card of fixed height with the topic's name, its decision so far, the gist of the previous turn if it was said on this topic, and the words of the turn being said (\textit{DG2}, \textit{DG4}).
The live words are shown a sentence at a time, or in runs of at most 14 words for a long sentence, and nine on screens shorter than 680\,px.
They are appended rather than revised word by word. 
A run is replaced as a whole when the voice reaches its next word, and when it starts mid-sentence its first three words are faded.
A caret in the speaker's colour follows the last word said (\textit{DG6}).
Because the card has a fixed height, a long turn can never push the earlier topics off the screen, which our first pilots of the stack showed it did (\textit{DG1}).

Above the card, every earlier topic keeps the slot it got when it was first left, alternating left and right, with the newest nearest the card (\textit{DG3}). Unlike panels per speaker, the side carries no meaning and the order reads from top to bottom.
The newest topics are shown as bubbles with their name and decision. When they no longer fit, the oldest become name-only pills and then clusters of four in order, opened in place with a tap.
The fit is worked out before the screen is drawn, so it is never seen settling.
Tapping a topic opens what each side said on it, and the words themselves one tap further.
Orange is the person's colour and blue the agent's, from the caret that follows the last word said up to the dot on each topic, which shows whose words it rests on and is split in two for a topic that rests on both.

\begin{figure*}[t]
    \centering
    \includegraphics[width=0.8\textwidth]{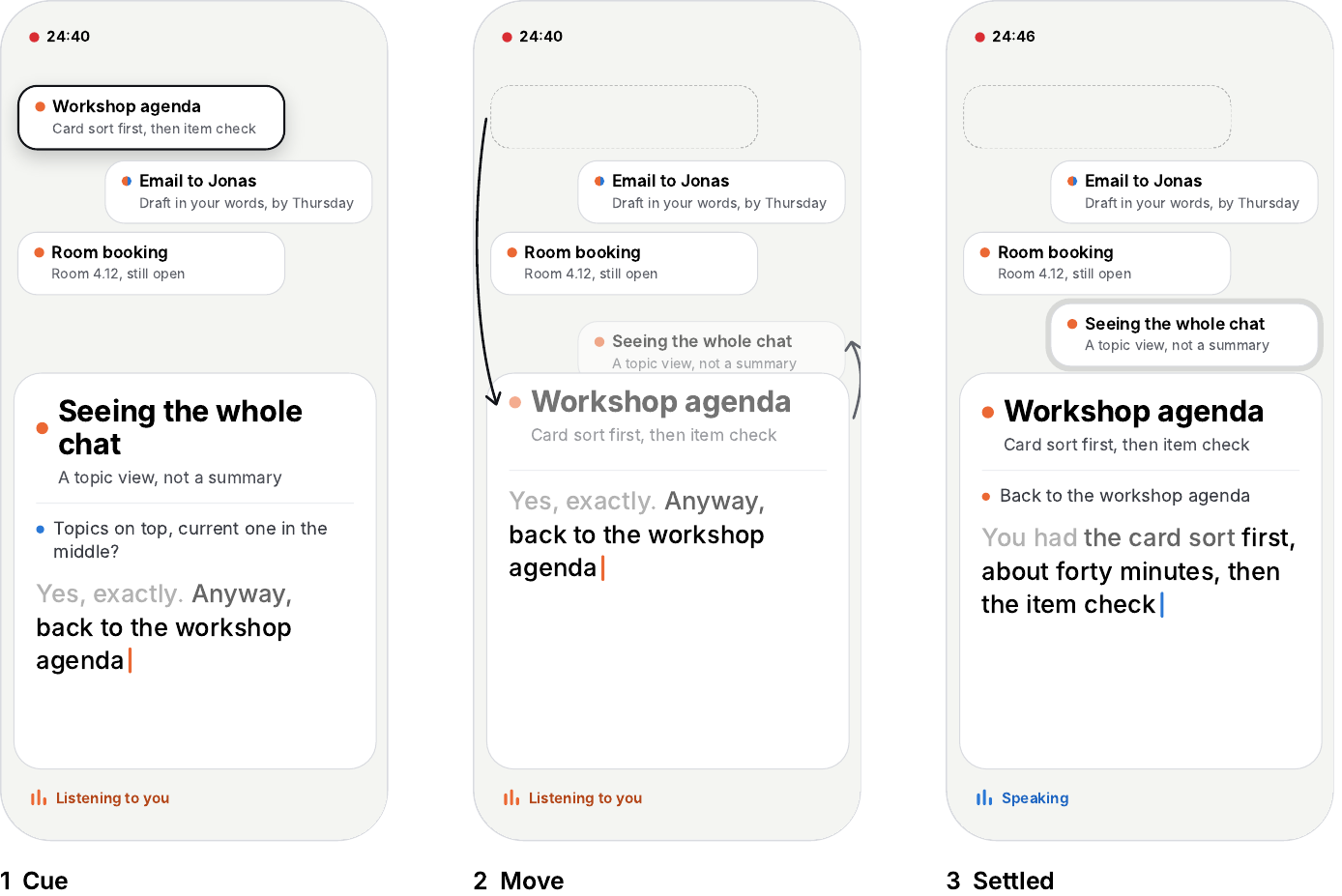}
    \caption{Returning to an earlier topic, the only animated move on the display (\textit{DG3}). 
    (1)~Cue: the titler names an earlier topic again, and its bubble lifts with a dark ring. 
    (2)~Move: the topic comes down into the card and the topic being left rises into a new slot. 
    A dashed outline keeps the old slot. 
    (3)~Settled: the decision line shows where the topic was left, and the topic just left keeps a soft ring for a glance that missed the move.}
    \Description{Three phone screens of the topic stack. In the first, the earlier topic ``Workshop agenda'' is outlined in black while the current topic is ``Seeing the whole chat''. In the second, arrows show ``Workshop agenda'' moving down into the large card and ``Seeing the whole chat'' moving up into a new slot, with a dashed outline left behind. In the third, ``Workshop agenda'' is the current topic and ``Seeing the whole chat'' sits in its new slot with a grey ring.}
    \label{fig:return}
\end{figure*}

A return to an earlier topic is the only animated move (\autoref{fig:return}).
The returning topic is first cued, lifting with a dark ring for 200\,ms.
It then comes down into the card in one eased move of 500\,ms without overshoot, while the topic being left rises into a slot of its own~\cite{chalbi2020commonfate,heer2007animatedtransitions}.
Its old slot stays as a dashed gap, so no other topic changes place.
Because a glance can miss the move~\cite{simons2005changeblindness}, the topic just left keeps a soft ring for 8\,s.
A new topic needs no cue: its name fades in and the topic left joins the earlier ones.
With reduced motion enabled, all moves are cuts and the rings remain.

\subsection{Study Apparatus}

The same app runs the study at \texttt{/study/session}, on phones only.
A setup lists the views and tasks to compare, and each new participant gets the least-filled combination of view and task.
The conversation can be finished after 4\,min and ends at 7\,min.
Between 2:15 and 3:00, at the first moment both sides have been quiet for 1.5\,s, or up to 30\,s past 3:00 if they never do, a game of tic-tac-toe against the phone takes over the screen for 60 to 75\,s. 
The phone plays a simple strategy with occasional random moves, so that the game holds attention without being trivial.
The app logs the end of the game and the onset of the person's first transcribed words after it, which give the resumption lag, and the phone's own voice detection as a check.
The procedure, the measures and the blind \ac{llm} judge are described with the evaluation (\autoref{sec:procedure}).
 
\section{Evaluation}
\label{sec:study}

In conversational AI systems, human agency moves into the conversation itself and the judgement of its outcome~\cite{wu2026humanagencyintheageofconversationalai}.
We evaluate Souffleur on how people follow the conversation (RQ1), how they resume it after an interruption (RQ2), and what they take away from it (RQ3).
Both studies share one procedure (\autoref{sec:procedure}).
A calibration study with nine participants tested this procedure and led to four adjustments (\autoref{sec:calibration}).
The main study then compared Souffleur with a conventional chat between subjects, with 48 participants (\autoref{sec:main-study}).

\subsection{Study Procedure}
\label{sec:procedure}

Participants were recruited on Prolific\footnote{\url{https://www.prolific.com}} and took part on their own smartphone without further instructions.
Each participant held one voice conversation with our evaluation system about a given task, while the phone showed the conversation in the display of their condition: 
a conventional chat (\condChat{}), or the presented implementation of Souffleur (\varSouffleur{}, \autoref{sec:implementation}).
The agent was a speech-to-speech model\footnote{\texttt{gpt-realtime-2.1} by OpenAI} with identical instructions in both conditions.

Both displays ran in the same app and received the same text, so they differed only in how they laid it out.
The \condChat{} display showed one message per turn, the person's turns as grey bubbles on the right and the agent's as plain text on the left, with the latest turn at the bottom. 
It was plainer than the illustration in \autoref{fig:teaser}a.
In both conditions, the agent's words appeared at the pace of its voice, the person's words appeared as they were transcribed, and the same status line said what the system was doing.
A slim bar above both displays showed the task, which participants could expand to reread the brief, and a Finish button.
The titler (\autoref{sec:implementation}) also ran invisibly during \condChat{} sessions, so that the topic of the first words after the interruption could be compared in both conditions. 
Decisions and gists were computed only for \varSouffleur{}.

After informed consent and a microphone check, participants read how the session works, played one practice round of the interrupting game, read the task, held the conversation and ended with a survey.

\subsubsection{Interruption: A Minute of Tic-Tac-Toe}

To study resumption (RQ2), we interrupted the conversation once with a game of tic-tac-toe.
The game took over the whole screen, so it interrupted the conversation and forced a context switch.
It appeared between 2:15 and 3:30 into the conversation, when both sides were silent.
There was no warning, because a warning lets people rehearse their goal and resume faster~\cite{trafton2003preparingtoresume}.
Any support for resuming therefore had to come from the display.
While the game ran, the microphone was muted and the agent paused. 
It was told not to recap and not to speak first when the conversation resumed.
Because longer and more demanding interruptions slow resumption~\cite{monk2008interruptionduration}, the game lasted between 60\,s and 75\,s before returning the participant to the conversation and to the display that had been interrupted.

\subsubsection{Measures}

\autoref{tab:study-measures} assigns dependent variables to research questions.
The primary measure is the \emph{resumption lag}, the time it takes to restart a task after an interruption~\cite{trafton2003preparingtoresume}. 
In our study, it is the time from the end of the interruption to the onset of the participant's speech.
The phone's own voice detection gave a second measure of this, but it ran about 1.3\,s behind the transcript and missed the first words in six sessions.
Therefore, we report it only as a check.
After the conversation, the participants summarised it, said where they had been before the game and how they had returned, and answered four ratings.
A blind \ac{llm} judge scored the resumption and the summaries from the transcripts.
For the resumption, it decided whether the participant's first words continued the topic before the interruption and whether the participant asked the agent where they had been, based on the conversation before and after the interruption.
For each summary, it listed the main content of the conversation, marked each as present, partly present or absent in the summary, and counted invented points that the conversation did not contain.
We measure recall with the share of key points in the summary, with partly present points counting half.
The judge also rated the summary's coverage, accuracy and specificity from 1 to 5.
For blinding, speakers were named only Person and Assistant, words naming an interface type were removed, and the items were judged in random order and with random identifiers.

\begin{table*}[t]
\caption{
    Dependent variables for each research question.
    Ratings are on 7-point scales with labelled ends.
}
\label{tab:study-measures}
\small
\begin{tabular}{@{}l>{\raggedright\arraybackslash}p{0.45\textwidth}>{\raggedright\arraybackslash}p{0.45\textwidth}@{}}
\toprule
RQ & Dependent variable & Collected \\
\midrule
RQ1 & \textbf{Ease of following}: ``How easy was it to follow the conversation?'' & Rating after the conversation \\
    & \textbf{Effort}: ``How much effort did keeping up with it take?'' & Rating after the conversation \\
    & \textbf{Looking at the screen}: ``How much did you look at the screen while you were talking?'' & Rating after the conversation \\
\midrule
RQ2 & \textbf{Resumption lag}: end of the game to the onset of the first transcribed words & App logging; phone's voice detection as a check \\
    & \textbf{Resumption accuracy}: the first words continue the topic open before the interruption (yes, no, unclear) & Blind \ac{llm} judge on the transcript around the interruption \\
    & \textbf{Ease of picking up}: ``After the game, how easy was it to pick up where you had left off?'' & Rating after the conversation (very hard to very easy) \\
    & \textbf{Resumption approach}: How they found their way back & Open question after the conversation \\
\midrule
RQ3 & \textbf{Visual support}: What on the screen helped or got in the way & Open question after the conversation \\
    & \textbf{Recall}: share of the conversation's key points in the summary; invented points; coverage, accuracy and specificity (1--5) & Free-recall summary after the conversation, scored by a blind \ac{llm} judge \\
\bottomrule
\end{tabular}
\end{table*}

\subsection{Calibration Study}
\label{sec:calibration}

The procedure, tasks and measures were themselves designed formatively.
Before the main study, we ran the complete procedure with a small sample and checked
(1)~the evaluation pipeline, from interaction tracking to the measures derived from it,
(2)~whether participants understood the tasks and the displays as intended,
and (3)~the study design itself.
The calibration crossed the two displays with two planning tasks, a weekend away for four people and a 40th birthday dinner for twelve, assigned in rotation.
Nine people enrolled. 
Seven held a conversation that was interrupted, and six of them completed the questionnaire.
Besides the study's own records, we read each session's interaction logs (\autoref{sec:implementation}).

The technical infrastructure held: every audio chunk was transcribed, and every interrupted conversation recorded the timing of its interruption.
We did not score the calibration summaries with the blind \ac{llm} judge, since its rubric for reflection tasks was itself written during the calibration.
However, four findings would have biased the measures, and each changed the procedure of \autoref{sec:procedure}:

\begin{itemize}
    \item \textbf{Speech during the interruption.}
    Several participants kept talking during the game, so a sentence begun during the interruption counted as their first words afterwards.
    The microphone is now muted for the interruption, and the first words are timed from the speech that follows a silence.

    \item \textbf{The agent cut off.}
    In two sessions the game began while the agent was speaking or composing a reply, and its sentence was never finished.
    The interruption now waits for silence for up to 30\,s past its deadline.

    \item \textbf{Technical reliability.}
    The study is meant to run on phones, and although the layout is comparable on larger screens, we restricted it to phones so that browser developer tools cannot be used.
    Because browsers machine-translated some screens, including the topics on the display, browser translation is disabled and the study is limited to English.

    \item \textbf{Task variety.}
    Both tasks were planning tasks, so the calibration could not show whether the effect of the display holds beyond one kind of conversation.
    The main study uses four tasks that cross planning and reflection with a given situation or the participant's own life (\autoref{sec:main-study}).
    
\end{itemize}

The calibration also showed how participants used the display.
Without any training in Souffleur, nobody tapped it during the conversation in either condition, and nobody opened the full wording that Souffleur keeps one tap away (DG4).
Asked how they had found their way back after the game, a participant in the \varSouffleur{} condition said they looked at the last topic, a participant in the \condChat{} condition at the agent's last message, and two others that they asked the agent.
The measures returned plausible values for resumption lags and for the other self-reported measures.
We did not compare the conditions on the calibration data, because this sample is too small~\cite{albers2018pilotdatastudysize}.

\subsection{Main Study}
\label{sec:main-study}

\subsubsection{Design}
We use a $2 \times 2 \times 2$ between-subjects design (\autoref{tab:study-design}).
The independent variable is the \emph{interface}, \condChat{} or \varSouffleur{}.
Two further task-specific variables test the impact of the interface beyond a single kind of conversation:
\emph{task type} contrasts planning (converging towards decisions) with reflection (open-ended, touching several topics).
\emph{task source} contrasts a situation given to the participant with one of their own choice, which may be more realistic to them.
In reflection tasks the agent was told to ask rather than advise, and in the participants' own tasks never to ask for identifying details.
Participants were assigned evenly to the eight cells, two interfaces crossed with four tasks.
A weekend away for four (planning, given) asked where to go, how all four get there and where to stay, for a trip with three friends, about 150 euros each and one person who does not drive.
Something the participant needed to plan in the coming weeks (planning, own) asked what has to happen and what to do first.
A described hard week and the participant's own past week (reflection, given and own) asked what went well, what did not and what to change.
Participants read the brief before the conversation and could reopen it during it.

\subsubsection{Participants}
We recruited 50 fluent English-speaking smartphone users on Prolific.
48 completed the study, 24 per interface and six per cell.
They were 18 to 59 years old, 31 were male and 17 female, living in 12 countries and for 38 of them English was not their first language.
The study took a median of 15 minutes, and participants received their compensation through Prolific.
With 24 participants per interface, a two-sided test at $\alpha = .05$ had 80\% power only for effects of about $d = 0.8$ or larger, and with 21 and 18 participants in the analysis of resumption, about $d = 0.9$~\cite{lakens2022samplesizejustification}.
We therefore treat the study as exploratory and report the comparisons as estimates rather than as tests of a fixed hypothesis.

\subsubsection{Analysis}
We excluded sessions that failed for technical reasons, as shown by the transcripts and interaction logs, for example sessions in which audio could not be transcribed or the browser repeatedly errored.
From the analysis of resumption we also excluded participants who did not continue to speak after the interruption.
We compared the log resumption lag with a linear model that included task type and task source as covariates, the ratings, summary length and judge scores with Mann--Whitney $U$ tests, and counts with Fisher's exact test.
As the analysis is exploratory, we did not correct for multiple comparisons.
The \ac{llm} judge was \texttt{GLM-5.3}, which scored every item twice.
We averaged the two scores per participant and counted a resumption as on topic only if both judgements said so.
To check that the scores did not depend on one model, a smaller open-weights model, \texttt{gemma3:12b}, scored the same items with the same rubric.
Both models agreed with themselves and with each other on the share of key points and on the resumption.
However, they did not agree on accuracy or on invented points, so we do not interpret these two scores.
We grouped the open answers on how participants found their way back into five categories, one category per answer.

\begin{table}[t]
\caption{
    Independent variables of the $2 \times 2 \times 2$ between-subjects design. 
    The interface is the variable of interest.
    The task variables test whether its effect holds across kinds of conversation.
    }
\label{tab:study-design}
\small
\begin{tabular}{@{}lll@{}}
\toprule
Variable & Level & Description \\
\midrule
Interface   & \condChat{} & One message per turn (baseline) \\
            & \varSouffleur{} & Topic stack \\
\midrule
Task type   & Planning & Converges on decisions \\
            & Reflection & Open-ended, few decisions \\
\midrule
Task source & Given & A predefined situation \\
            & Own & Self chosen scenario \\
\bottomrule
\end{tabular}
\end{table}

\subsection{Ethical Considerations}
\label{sec:ethics}

Under Danish law, this research does not require approval from an ethics committee because it is not a health research project and no health data were collected.
All participants were adults recruited on Prolific and gave informed consent in the app before the session.
Data were processed in accordance with the GDPR.

\section{Results}
\label{sec:results}

Of the 48 participants who completed the study, 46 provided complete data for all analyses (\condChat{} 22, \varSouffleur{} 24).
Seven of them finished within 15\,s after the interruption without continuing to speak.
Therefore, they had no resumption lag, which leaves 39 participants (\condChat{} 21, \varSouffleur{} 18) for the analysis of resumption.
\autoref{tab:results} summarises the measures by interface.

\begin{table*}[t]
\caption{
    Measures by interface. 
    Ratings on 7-point scales and judge scores as mean (SD).
    The resumption lag is the time from the end of the interruption to the onset of the first transcribed words. 
    $p$: linear model on the log lag with task type and task source as covariates for the lags, Mann--Whitney $U$ for ratings, summary length and judge scores, Fisher's exact test for counts.
    }
\label{tab:results}
\small
\begin{tabular}{@{}llccr@{}}
\toprule
RQ & Measure & \condChat{} & \varSouffleur{} & $p$ \\
\midrule
RQ1 & Ease of following (very hard to very easy) & 5.86 (1.28) & 5.92 (1.21) & .95 \\
    & Effort (very little to a great deal) & 3.77 (1.82) & 3.88 (2.07) & .90 \\
    & Looking at the screen (hardly at all to most of the time) & 5.45 (1.10) & 5.04 (1.52) & .36 \\
\midrule
RQ2 & Resumption lag, median [IQR] (s) & 2.89 [2.02, 4.85] & 3.27 [2.12, 6.10] & .71 \\
    & Resumption lag from the phone's voice detection, median (s) & 3.86 & 5.78 & .19 \\
    & First words on the topic suspended before the interruption & 20 of 21 & 18 of 18 & \\
    & Resumed on the suspended topic, \ac{llm} judge & 18 of 21 & 15 of 18 & 1.00 \\
    & Ease of picking up (very hard to very easy) & 5.77 (1.34) & 5.21 (1.72) & .27 \\
    & Finished within 15\,s of the interruption's end & 1 of 22 & 6 of 24 & .10 \\
\midrule
RQ3 & Summary length, median [IQR] (characters) & 165 [100, 212] & 93 [74, 147] & .09 \\
    & Key points in the summary, \ac{llm} judge (\%) & 27 (14) & 22 (14) & .40 \\
    & Coverage of the summary, \ac{llm} judge (1--5) & 1.68 (0.63) & 1.46 (0.51) & .23 \\
    & Specificity of the summary, \ac{llm} judge (1--5) & 1.57 (0.74) & 1.46 (0.72) & .57 \\
\midrule
    & $n$ (ratings, summaries; resumption lag) & 22; 21 & 24; 18 & \\
\bottomrule
\end{tabular}
\end{table*}

\subsection{Following the Conversation (RQ1)}

Participants in the two conditions rated the conversation as similarly easy to follow (medians of 6 out of 7), reported similar effort, and looked at the screen for much of the conversation in both (medians 5.5 and 5). 
None of the differences was reliable (\autoref{tab:results}).
\subsection{Resuming after the Interruption (RQ2)}

The interruption came a median of 161\,s into the conversation (range 135--210\,s) and lasted 60 to 74\,s.
Participants resumed after a median of 2.9\,s with \condChat{} and 3.3\,s with \varSouffleur{}.
In the linear model, the lag with \varSouffleur{} was 1.10 times that with \condChat{} (95\% CI 0.65 to 1.88, $p = .71$).
The task type and task source did not have a reliable effect.
The lag from the phone's voice detection, a check on the transcript measure, showed the same direction.
Almost all participants resumed on the topic they had left: in 38 of 39 conversations, the first words after the interruption were filed under the same topic as the last words before it.
Because the titler keeps the current title unless the subject changes, this measure leans towards continuity.
The blind \ac{llm} judge, reading the transcripts, was stricter and found that the first words after the interruption continued the suspended topic for 18 of 21 participants with \condChat{} and 15 of 18 with \varSouffleur{}.
No participant asked the agent where they had been.
Ratings of how easy it was to pick up where they had left off did not differ between the interfaces.

Two behaviours around the resumption differed from what the design intended.
First, seven participants ended the conversation instead of resuming it.
One participant explained, ``when I got back in, I had the option available to finish the conversation, so I ended it''.
Second, in four conversations the agent answered the participant's first word after the interruption as if it were a complete turn, within about two seconds.
The agent joined a resumption that the study meant to leave to the participant.

\autoref{tab:way-back} groups participants' own accounts of how they found their way back.
Participants in the \condChat{} condition more often named the screen (6 of 22, against 2 of 24 with \varSouffleur{}), typically by rereading the last messages (``I looked at the chat transcript on the screen to see what was said last and continued from there'').
Participants in the \varSouffleur{} condition more often described resuming from memory, for example ``I remembered where I stopped'' or ``I took the time when I'm playing to think and remember''.
In both conditions, three participants described the agent as taking the lead, and some found the moment of return unclear: ``it was just paused, I didn't know what to do''.

\begin{table}[t]
\caption{
    The different categories group participants' answers about how they found their way back into the conversation after the interruption.
    }
\label{tab:way-back}
\small
\begin{tabular}{@{}lcc@{}}
\toprule
Way back & \condChat{} & \varSouffleur{} \\
\midrule
Looked at the screen & 6 & 2 \\
From memory or own next question & 5 & 7 \\
The agent took the lead & 3 & 3 \\
Ended the conversation & 1 & 3 \\
No strategy named (``easy'', ``automatically'') & 7 & 9 \\
\midrule
Total & 22 & 24 \\
\bottomrule
\end{tabular}
\end{table}

\subsection{What Participants Took Away (RQ3)}

Asked what on the screen had helped or got in the way, participants in the \varSouffleur{} condition named its structure: ``the bullet-point-style headlights'' [sic], ``boxes highlighting main points of the conversation'', and ``it would list the topics we were discussing, so that helped a lot''.
One found it ``a bit confusing'', with ``too much information offered by the assistant without much context''.
In the \condChat{} condition, several comments concerned the speech recognition rather than the content.
Five participants mentioned its recognition errors or its delay (``the text only seemed to appear significantly after I heard the same words''), while three found the word-for-word text helpful (``the conversation appearing word for word helped me remember the context'').

Participants with \varSouffleur{} wrote shorter summaries than those with \condChat{} (a median of 93 against 165 characters, or about 20 against 30 words).
Asked where they had been just before the game, 4 of 22 participants with \condChat{} and 2 of 24 with \varSouffleur{} said they did not remember.
The summaries captured little of the conversation with either interface.
The judge identified about seven key points per conversation (range 5 to 8), of which the summaries contained 27\% with \condChat{} and 22\% with \varSouffleur{}.
The judge's ratings of coverage (medians 1.75 and 1.25 out of 5) and specificity (medians of 1 in both conditions) were low and did not differ between the interfaces either.
The difference followed the length of the summaries:
longer summaries contained more key points, and with the length in the model, the interface made no difference.

\section{Discussion}
\label{sec:discussion}

We asked how a glanceable topic display compares with the conventional chat display of stacked messages in how people follow a voice conversation on a phone (RQ1), resume it after an interruption (RQ2) and take away what it was about (RQ3).
The chat display is the most familiar way to hold such conversations, which makes it a strong baseline.
For none of the three did \varSouffleur{} perform reliably better or worse than the transcript.
Given the effects our study could detect, this rules out only large differences in either direction.

\subsection{Analytical Evaluation}

Beyond the comparison, we assess Souffleur against the systems criteria of Olsen~\cite{olsen2007evaluation} that apply to a display rather than a toolkit.
\emph{Expressive match}: topics and decision lines use the conversation's own concrete terms, such as ``Room 4.12, still open'', and participants described them as ``the topics we were discussing'' without any training.
\emph{Generality}: the same pipeline served planning and reflection, given and own tasks without changes, and folded every conversation into topics.
\emph{Reducing the skill barrier}: participants in 12 countries, most of them speaking English as a second language, used it on their own phones without instruction and rated it about as easy to follow as the familiar chat.
\emph{Scale} is designed but remains untested: the conversations had a median of two topics (range 1--5), too few to exercise the folding into pills and clusters.

\subsection{No Detectable Difference in a Short Conversation}

We suppose that resumption was easy for the given tasks in general.
The transcript and the judge agree that almost all participants continued the topic that they had suspended.
A goal that is suspended for a short time and was active just before is still accessible from memory~\cite{altmann2002memoryforgoals}, and the cost of an interruption grows with its length and demand~\cite{monk2008interruptionduration}.
Our interruption was short, and the conversation before it was only two to three minutes, with few topics.
There was little to forget and little for the display to restore.
Sixteen of 46 participants named no strategy for finding their way back, and only eight named the screen.
A display that holds the suspended goal can only shorten a resumption that memory alone makes slow, which our setting did not produce.
Moreover, participants held the conversation without a concurrent task and could look at the screen freely, so the display never had to work in the time of a glance.

\textbf{The transcript is a strong baseline for resuming.}
The last lines of a transcript show the words spoken just before the interruption, which is the cue people rehearse and look for to resume a task~\cite{trafton2003preparingtoresume}.
Participants with \condChat{} used it this way, and six of them explicitly mentioned rereading the last messages to continue.
When the game ended, Souffleur showed the topic, its decision line, a gist of the turn before, and only the end of the last turn, its final sentence or at most 14 words, which serves an overview of many topics better than a return to the most recent exchange.
This suggests that a compacted display should keep the most recent exchange readable in full, a requirement that also applies to other visualisations of conversational content~\cite{chandrasegaran2019talktraces,chen2025meetmap,coscia2025ongoal}.

\textbf{We found no effect of the display on what participants took away.}
Saying information aloud already supports remembering it~\cite{macleod2010theproductioneffect}, and if it was generated the remembering is even stronger~\cite{slamecka1978thegenerationeffect}.
In our study, the screen did not detectably add to this.
Recall was low in both conditions and mostly contained the topics in general terms.
Participants with \varSouffleur{} wrote shorter summaries, which may mean that they relied on the display rather than on their own account of the conversation.
Telling these apart would require research that also includes metacognitive dimensions.

\subsection{Implications for Glanceable Voice Displays}

Our results show no detectable advantage or disadvantage of a structured display for short voice conversations.
Glanceable displays are meant to inform at a glance without demanding attention~\cite{matthews2006glanceable}.
By that standard, Souffleur cost no more effort or attention than the transcript, while showing the conversation in a fraction of the words.
Its intended advantage, an overview of where a conversation stands, should matter for longer conversations with many topics, longer or more demanding interruptions, and returns after hours or days rather than a minute.
These settings and a longitudinal study are the next test for our design goals.

\subsection{Limitations}

With 22 and 24 participants in the analyses, the study could detect only large differences.
Our results are therefore not evidence that the displays are equivalent.
Most importantly, participants held the conversation without a concurrent task and looked at the screen for much of it, so the situations that motivate this work were not tested.
The participants were recruited online and held one conversation of a few minutes on their own phone with a minute of interruption, but interruptions in daily use are longer, less predictable and even physical, such as crossing a street or talking to someone else.
The tasks were set by us rather than the participants' own obligations, even when participants planned or reflected on their own lives.
Most participants spoke English as a second language, which may have affected the speech recognition and the interaction.
Also, seven participants ended the conversation right after the game, six of them with \varSouffleur{}, and in four conversations the agent answered the participant's first word.
Both reduce the number of resumptions that reflect the display alone, and the first may bias the comparison in favour of \varSouffleur{}.
Resumption accuracy and recall were scored by \ac{llm} judges.
We reviewed the judges' output and the chosen models' benchmarks, but did not explicitly evaluate all answers.

Our design goals are not exhaustive, and other designs that meet them, such as a topic map, may perform differently.
As we describe in \autoref{sec:formative-activities}, we iterated on several prototypes and consider the presented design the one that best met our design goals, but we did not evaluate the other designs in a controlled study.

\subsection{Future Work}

Beyond longer conversations and interruptions, three directions follow from our work.
First, we focused on the phone because it is the device people carry into the situations that motivate this work, such as walking, cooking or commuting.
We encourage research in this direction to keep its focus on smaller screens.
Second, the display could move from showing the conversation to acting on it: topics that can be reopened, decisions that become tasks, and integration with the tools in which plans are carried out.
Third, it could also serve conversations with other people, in group meetings or as a side display in online meetings, though its provenance signals would need to cover more than two voices.
We encourage other designs to challenge ours on the design goals we derived.

\section{Conclusion}

Voice conversations with \ac{ai} increasingly happen on the phone, in moments where the screen can get no more than a glance.
We presented Souffleur, a voice-first thinking partner whose display folds the conversation into a stack of topics, each with the decision reached on it, following six design goals derived from formative work.
In a between-subjects study with 48 participants on their own phones, \varSouffleur{} and a conventional chat transcript did not differ reliably in the ease of following the conversation, the way of resumption or the recall.
The value of a glanceable overview can be tested further in longer conversations with longer interruptions.
We hope that our design goals, our study procedure and these results help others build and test such displays, and extend them from showing a conversation to acting on it.

\begin{acks}
    We especially thank Johannes Ellemose for advising and supporting the study.
    This work was supported partly by Villum Investigator grant VL-54492 by Villum Fonden.
    Any opinions, findings, and conclusions expressed in this material are those of the authors and do not necessarily reflect the views of the funding agency.
\end{acks}

\section*{Use of Generative AI}
\label{sec:ai-disclosure}

OpenAI's realtime speech-to-speech model held the conversation with participants, further language models wrote the topic headlines and decision lines on the display, and a blind \ac{llm} judge running \texttt{GLM-5.3}, cross-checked by \texttt{gemma3:12b}, scored the resumptions and summaries.
The system was implemented with the help of agentic coding tools (Claude Code with Claude Opus~5.5 and Claude Sonnet~5.5; OpenCode with \texttt{GLM-5.3}).
Claude Opus~5.5 helped write the analysis scripts and summarise the calibration logs.
The illustrations were drafted with Claude (Anthropic) on a design canvas and edited by the authors with its editing tools or exported into other design tools, such as Figma.

We used Claude Opus~5.5 (Anthropic), the self-hosted \texttt{GLM-5.3} and Consensus (consensus.app, accessed August--October 2026), an AI search engine without specific \ac{llm} models indicated, to brainstorm and discuss design and study decisions, to search for and summarise related literature, and to suggest edits to the text.
Project meetings were transcribed automatically with Gemini in Google Meet.
All decisions were made by the authors, every cited source was checked against its publisher record, and all text was revised by the authors, who take full responsibility for the content of this paper.

\bibliographystyle{ACM-Reference-Format}
\bibliography{voice-ai}


\begin{thebibliography}{66}

%%% the Web address.

\ifx \showCODEN    \undefined \def \showCODEN     #1{\unskip}     \fi
\ifx \showISBNx    \undefined \def \showISBNx     #1{\unskip}     \fi
\ifx \showISBNxiii \undefined \def \showISBNxiii  #1{\unskip}     \fi
\ifx \showISSN     \undefined \def \showISSN      #1{\unskip}     \fi
\ifx \showLCCN     \undefined \def \showLCCN      #1{\unskip}     \fi
\ifx \shownote     \undefined \def \shownote      #1{#1}          \fi
\ifx \showarticletitle \undefined \def \showarticletitle #1{#1}   \fi
\ifx \showURL      \undefined \def \showURL       {\relax}        \fi
% invisible to TeX
\providecommand\bibfield[2]{#2}
\providecommand\bibinfo[2]{#2}
\providecommand\natexlab[1]{#1}
\providecommand\showeprint[2][]{arXiv:#2}

\bibitem[Albers and Lakens(2018)]%
        {albers2018pilotdatastudysize}
\bibfield{author}{\bibinfo{person}{Casper Albers} {and}
  \bibinfo{person}{Daniël Lakens}.} \bibinfo{year}{2018}\natexlab{}.
\newblock \showarticletitle{When power analyses based on pilot data are biased:
  Inaccurate effect size estimators and follow-up bias}.
\newblock \bibinfo{journal}{\emph{Journal of Experimental Social Psychology}}
  \bibinfo{volume}{74} (\bibinfo{year}{2018}), \bibinfo{pages}{187--195}.
\newblock
\showISSN{0022-1031}
\href{https://doi.org/10.1016/j.jesp.2017.09.004}{doi:\nolinkurl{10.1016/j.jesp.2017.09.004}}


\bibitem[Altmann and Trafton(2002)]%
        {altmann2002memoryforgoals}
\bibfield{author}{\bibinfo{person}{Erik~M. Altmann} {and}
  \bibinfo{person}{J.~Gregory Trafton}.} \bibinfo{year}{2002}\natexlab{}.
\newblock \showarticletitle{Memory for goals: an activation-based model}.
\newblock \bibinfo{journal}{\emph{Cognitive Science}} \bibinfo{volume}{26},
  \bibinfo{number}{1} (\bibinfo{year}{2002}), \bibinfo{pages}{39--83}.
\newblock
\href{https://doi.org/10.1207/s15516709cog2601\_2}{doi:\nolinkurl{10.1207/s15516709cog2601\_2}}


\bibitem[Baghdadi et~al\mbox{.}(2025)]%
        {baghdadi2025richhud}
\bibfield{author}{\bibinfo{person}{Mahmoud Baghdadi}, \bibinfo{person}{Dilara
  Samad-Zada}, {and} \bibinfo{person}{Achim Ebert}.}
  \bibinfo{year}{2025}\natexlab{}.
\newblock \showarticletitle{Evaluating Rich Visual Feedback on Head-Up Displays
  for In-Vehicle Voice Assistants: A User Study}.
\newblock \bibinfo{journal}{\emph{Multimodal Technologies and Interaction}}
  \bibinfo{volume}{9}, \bibinfo{number}{11} (\bibinfo{date}{Nov.}
  \bibinfo{year}{2025}), \bibinfo{pages}{114}.
\newblock
\showISSN{2414-4088}
\href{https://doi.org/10.3390/mti9110114}{doi:\nolinkurl{10.3390/mti9110114}}


\bibitem[Bartram et~al\mbox{.}(2003)]%
        {bartram2003moticons}
\bibfield{author}{\bibinfo{person}{Lyn Bartram}, \bibinfo{person}{Colin Ware},
  {and} \bibinfo{person}{Tom Calvert}.} \bibinfo{year}{2003}\natexlab{}.
\newblock \showarticletitle{Moticons: Detection, Distraction and Task}.
\newblock \bibinfo{journal}{\emph{International Journal of Human-Computer
  Studies}} \bibinfo{volume}{58}, \bibinfo{number}{5} (\bibinfo{year}{2003}),
  \bibinfo{pages}{515--545}.
\newblock
\href{https://doi.org/10.1016/S1071-5819(03)00021-1}{doi:\nolinkurl{10.1016/S1071-5819(03)00021-1}}


\bibitem[Bergstrom and Karahalios(2007)]%
        {bergstrom2007conversationclock}
\bibfield{author}{\bibinfo{person}{Tony Bergstrom} {and}
  \bibinfo{person}{Karrie Karahalios}.} \bibinfo{year}{2007}\natexlab{}.
\newblock \showarticletitle{Conversation Clock: Visualizing audio patterns in
  co-located groups}. In \bibinfo{booktitle}{\emph{2007 40th Annual Hawaii
  International Conference on System Sciences (HICSS'07)}}.
  \bibinfo{publisher}{IEEE}, \bibinfo{pages}{78--78}.
\newblock
\href{https://doi.org/10.1109/HICSS.2007.151}{doi:\nolinkurl{10.1109/HICSS.2007.151}}


\bibitem[Cha et~al\mbox{.}(2020)]%
        {cha2020opportunemoments}
\bibfield{author}{\bibinfo{person}{Narae Cha}, \bibinfo{person}{Auk Kim},
  \bibinfo{person}{Cheul~Young Park}, \bibinfo{person}{Soowon Kang},
  \bibinfo{person}{Mingyu Park}, \bibinfo{person}{Jae-Gil Lee},
  \bibinfo{person}{Sangsu Lee}, {and} \bibinfo{person}{Uichin Lee}.}
  \bibinfo{year}{2020}\natexlab{}.
\newblock \showarticletitle{Hello There! Is Now a Good Time to Talk? Opportune
  Moments for Proactive Interactions with Smart Speakers}.
\newblock \bibinfo{journal}{\emph{Proc. ACM Interact. Mob. Wearable Ubiquitous
  Technol.}} \bibinfo{volume}{4}, \bibinfo{number}{3}, Article
  \bibinfo{articleno}{74} (\bibinfo{date}{Sept.} \bibinfo{year}{2020}),
  \bibinfo{numpages}{28}~pages.
\newblock
\href{https://doi.org/10.1145/3411810}{doi:\nolinkurl{10.1145/3411810}}


\bibitem[Chalbi et~al\mbox{.}(2020)]%
        {chalbi2020commonfate}
\bibfield{author}{\bibinfo{person}{Amira Chalbi}, \bibinfo{person}{Jacob
  Ritchie}, \bibinfo{person}{Deokgun Park}, \bibinfo{person}{Jungu Choi},
  \bibinfo{person}{Nicolas Roussel}, \bibinfo{person}{Niklas Elmqvist}, {and}
  \bibinfo{person}{Fanny Chevalier}.} \bibinfo{year}{2020}\natexlab{}.
\newblock \showarticletitle{Common Fate for Animated Transitions in
  Visualization}.
\newblock \bibinfo{journal}{\emph{IEEE Transactions on Visualization and
  Computer Graphics}} \bibinfo{volume}{26}, \bibinfo{number}{1}
  (\bibinfo{date}{Jan.} \bibinfo{year}{2020}), \bibinfo{pages}{386--396}.
\newblock
\showISSN{1077-2626}
\href{https://doi.org/10.1109/TVCG.2019.2934288}{doi:\nolinkurl{10.1109/TVCG.2019.2934288}}


\bibitem[Chandrasegaran et~al\mbox{.}(2019)]%
        {chandrasegaran2019talktraces}
\bibfield{author}{\bibinfo{person}{Senthil Chandrasegaran},
  \bibinfo{person}{Chris Bryan}, \bibinfo{person}{Hidekazu Shidara},
  \bibinfo{person}{Tung-Yen Chuang}, {and} \bibinfo{person}{Kwan-Liu Ma}.}
  \bibinfo{year}{2019}\natexlab{}.
\newblock \showarticletitle{TalkTraces: Real-Time Capture and Visualization of
  Verbal Content in Meetings}. In \bibinfo{booktitle}{\emph{Proceedings of the
  2019 CHI Conference on Human Factors in Computing Systems}} (Glasgow,
  Scotland Uk) \emph{(\bibinfo{series}{CHI '19})}.
  \bibinfo{publisher}{Association for Computing Machinery},
  \bibinfo{address}{New York, NY, USA}, \bibinfo{pages}{1–14}.
\newblock
\showISBNx{9781450359702}
\href{https://doi.org/10.1145/3290605.3300807}{doi:\nolinkurl{10.1145/3290605.3300807}}


\bibitem[Chen et~al\mbox{.}(2023)]%
        {chen2023screenornoscreen}
\bibfield{author}{\bibinfo{person}{Chen Chen}, \bibinfo{person}{Ella~T Lifset},
  \bibinfo{person}{Yichen Han}, \bibinfo{person}{Arkajyoti Roy},
  \bibinfo{person}{Michael Hogarth}, \bibinfo{person}{Alison~A Moore},
  \bibinfo{person}{Emilia Farcas}, {and} \bibinfo{person}{Nadir Weibel}.}
  \bibinfo{year}{2023}\natexlab{}.
\newblock \showarticletitle{Screen or No Screen? Lessons Learnt from a
  Real-World Deployment Study of Using Voice Assistants With and Without
  Touchscreen for Older Adults}. In \bibinfo{booktitle}{\emph{Proceedings of
  the 25th International ACM SIGACCESS Conference on Computers and
  Accessibility}} (New York, NY, USA) \emph{(\bibinfo{series}{ASSETS '23})}.
  \bibinfo{publisher}{Association for Computing Machinery},
  \bibinfo{address}{New York, NY, USA}, Article \bibinfo{articleno}{52},
  \bibinfo{numpages}{21}~pages.
\newblock
\showISBNx{9798400702204}
\href{https://doi.org/10.1145/3597638.3608378}{doi:\nolinkurl{10.1145/3597638.3608378}}


\bibitem[Chen et~al\mbox{.}(2025)]%
        {chen2025meetmap}
\bibfield{author}{\bibinfo{person}{Xinyue Chen}, \bibinfo{person}{Nathan Yap},
  \bibinfo{person}{Xinyi Lu}, \bibinfo{person}{Aylin Gunal}, {and}
  \bibinfo{person}{Xu Wang}.} \bibinfo{year}{2025}\natexlab{}.
\newblock \showarticletitle{MeetMap: Real-Time Collaborative Dialogue Mapping
  with LLMs in Online Meetings}.
\newblock \bibinfo{journal}{\emph{Proc. ACM Hum.-Comput. Interact.}}
  \bibinfo{volume}{9}, \bibinfo{number}{2}, Article
  \bibinfo{articleno}{CSCW132} (\bibinfo{date}{May} \bibinfo{year}{2025}),
  \bibinfo{numpages}{35}~pages.
\newblock
\href{https://doi.org/10.1145/3711030}{doi:\nolinkurl{10.1145/3711030}}


\bibitem[Cockburn et~al\mbox{.}(2009)]%
        {cockburn2009overviewdetail}
\bibfield{author}{\bibinfo{person}{Andy Cockburn}, \bibinfo{person}{Amy
  Karlson}, {and} \bibinfo{person}{Benjamin~B. Bederson}.}
  \bibinfo{year}{2009}\natexlab{}.
\newblock \showarticletitle{A review of overview+detail, zooming, and
  focus+context interfaces}.
\newblock \bibinfo{journal}{\emph{ACM Comput. Surv.}} \bibinfo{volume}{41},
  \bibinfo{number}{1}, Article \bibinfo{articleno}{2} (\bibinfo{date}{Jan.}
  \bibinfo{year}{2009}), \bibinfo{numpages}{31}~pages.
\newblock
\showISSN{0360-0300}
\href{https://doi.org/10.1145/1456650.1456652}{doi:\nolinkurl{10.1145/1456650.1456652}}


\bibitem[Coscia et~al\mbox{.}(2025)]%
        {coscia2025ongoal}
\bibfield{author}{\bibinfo{person}{Adam~J Coscia}, \bibinfo{person}{Shunan
  Guo}, \bibinfo{person}{Eunyee Koh}, {and} \bibinfo{person}{Alex Endert}.}
  \bibinfo{year}{2025}\natexlab{}.
\newblock \showarticletitle{OnGoal: Tracking and Visualizing Conversational
  Goals in Multi-Turn Dialogue with Large Language Models}. In
  \bibinfo{booktitle}{\emph{Proceedings of the 38th Annual ACM Symposium on
  User Interface Software and Technology}} (Busan, Republic of Korea)
  \emph{(\bibinfo{series}{UIST '25})}. \bibinfo{publisher}{Association for
  Computing Machinery}, \bibinfo{address}{New York, NY, USA}, Article
  \bibinfo{articleno}{208}, \bibinfo{numpages}{18}~pages.
\newblock
\showISBNx{9798400720376}
\href{https://doi.org/10.1145/3746059.3747746}{doi:\nolinkurl{10.1145/3746059.3747746}}


\bibitem[Dhanoa et~al\mbox{.}(2026)]%
        {dhanoa2026heydashboard}
\bibfield{author}{\bibinfo{person}{Vaishali Dhanoa}, \bibinfo{person}{Gabriela
  Molina~Le{\'o}n}, \bibinfo{person}{Eve Hoggan}, \bibinfo{person}{Eduard
  Gr{\"o}ller}, \bibinfo{person}{Marc Streit}, {and} \bibinfo{person}{Niklas
  Elmqvist}.} \bibinfo{year}{2026}\natexlab{}.
\newblock \showarticletitle{"Hey Dashboard!": Supporting Voice, Text, and
  Pointing Modalities in Dashboard Onboarding using Large Language Models}. In
  \bibinfo{booktitle}{\emph{Proceedings of the 2026 CHI Conference on Human
  Factors in Computing Systems}} (Barcelona, Spain) \emph{(\bibinfo{series}{CHI
  '26})}. \bibinfo{publisher}{Association for Computing Machinery},
  \bibinfo{address}{New York, NY, USA}, Article \bibinfo{articleno}{1},
  \bibinfo{numpages}{15}~pages.
\newblock
\showISBNx{9798400722783}
\href{https://doi.org/10.1145/3772318.3791766}{doi:\nolinkurl{10.1145/3772318.3791766}}


\bibitem[Draxler et~al\mbox{.}(2024)]%
        {draxler2024aighostwritereffect}
\bibfield{author}{\bibinfo{person}{Fiona Draxler}, \bibinfo{person}{Anna
  Werner}, \bibinfo{person}{Florian Lehmann}, \bibinfo{person}{Matthias Hoppe},
  \bibinfo{person}{Albrecht Schmidt}, \bibinfo{person}{Daniel Buschek}, {and}
  \bibinfo{person}{Robin Welsch}.} \bibinfo{year}{2024}\natexlab{}.
\newblock \showarticletitle{The AI Ghostwriter Effect: When Users do not
  Perceive Ownership of AI-Generated Text but Self-Declare as Authors}.
\newblock \bibinfo{journal}{\emph{ACM Trans. Comput.-Hum. Interact.}}
  \bibinfo{volume}{31}, \bibinfo{number}{2}, Article \bibinfo{articleno}{25}
  (\bibinfo{date}{Feb.} \bibinfo{year}{2024}), \bibinfo{numpages}{40}~pages.
\newblock
\showISSN{1073-0516}
\href{https://doi.org/10.1145/3637875}{doi:\nolinkurl{10.1145/3637875}}


\bibitem[El~Alaoui et~al\mbox{.}(2025)]%
        {elalaoui2025stepwrite}
\bibfield{author}{\bibinfo{person}{Hamza El~Alaoui}, \bibinfo{person}{Atieh
  Taheri}, \bibinfo{person}{Yi-Hao Peng}, {and} \bibinfo{person}{Jeffrey~P
  Bigham}.} \bibinfo{year}{2025}\natexlab{}.
\newblock \showarticletitle{StepWrite: Adaptive Planning for Speech-Driven Text
  Generation}. In \bibinfo{booktitle}{\emph{Proceedings of the 38th Annual ACM
  Symposium on User Interface Software and Technology}} (Busan, Republic of
  Korea) \emph{(\bibinfo{series}{UIST '25})}. \bibinfo{publisher}{Association
  for Computing Machinery}, \bibinfo{address}{New York, NY, USA}, Article
  \bibinfo{articleno}{86}, \bibinfo{numpages}{30}~pages.
\newblock
\showISBNx{9798400720376}
\href{https://doi.org/10.1145/3746059.3747610}{doi:\nolinkurl{10.1145/3746059.3747610}}


\bibitem[Endsley(1995)]%
        {endsley1995situationawareness}
\bibfield{author}{\bibinfo{person}{Mica~R. Endsley}.}
  \bibinfo{year}{1995}\natexlab{}.
\newblock \showarticletitle{Toward a Theory of Situation Awareness in Dynamic
  Systems}.
\newblock \bibinfo{journal}{\emph{Human Factors}} \bibinfo{volume}{37},
  \bibinfo{number}{1} (\bibinfo{year}{1995}), \bibinfo{pages}{32--64}.
\newblock
\href{https://doi.org/10.1518/001872095779049543}{doi:\nolinkurl{10.1518/001872095779049543}}


\bibitem[Findlater and McGrenere(2004)]%
        {findlater2004menus}
\bibfield{author}{\bibinfo{person}{Leah Findlater} {and}
  \bibinfo{person}{Joanna McGrenere}.} \bibinfo{year}{2004}\natexlab{}.
\newblock \showarticletitle{A comparison of static, adaptive, and adaptable
  menus}. In \bibinfo{booktitle}{\emph{Proceedings of the SIGCHI Conference on
  Human Factors in Computing Systems}} (Vienna, Austria)
  \emph{(\bibinfo{series}{CHI '04})}. \bibinfo{publisher}{Association for
  Computing Machinery}, \bibinfo{address}{New York, NY, USA},
  \bibinfo{pages}{89–96}.
\newblock
\showISBNx{1581137028}
\href{https://doi.org/10.1145/985692.985704}{doi:\nolinkurl{10.1145/985692.985704}}


\bibitem[Forrin and MacLeod(2017)]%
        {forrin2018hearingoneself}
\bibfield{author}{\bibinfo{person}{Noah~D. Forrin} {and}
  \bibinfo{person}{Colin~M. MacLeod}.} \bibinfo{year}{2017}\natexlab{}.
\newblock \showarticletitle{This time it's personal: the memory benefit of
  hearing oneself}.
\newblock \bibinfo{journal}{\emph{Memory}} \bibinfo{volume}{26},
  \bibinfo{number}{4} (\bibinfo{date}{Oct.} \bibinfo{year}{2017}),
  \bibinfo{pages}{574--579}.
\newblock
\showISSN{1464-0686}
\href{https://doi.org/10.1080/09658211.2017.1383434}{doi:\nolinkurl{10.1080/09658211.2017.1383434}}


\bibitem[Gajos et~al\mbox{.}(2008)]%
        {gajos2008predictability}
\bibfield{author}{\bibinfo{person}{Krzysztof~Z. Gajos},
  \bibinfo{person}{Katherine Everitt}, \bibinfo{person}{Desney~S. Tan},
  \bibinfo{person}{Mary Czerwinski}, {and} \bibinfo{person}{Daniel~S. Weld}.}
  \bibinfo{year}{2008}\natexlab{}.
\newblock \showarticletitle{Predictability and accuracy in adaptive user
  interfaces}. In \bibinfo{booktitle}{\emph{Proceedings of the SIGCHI
  Conference on Human Factors in Computing Systems}} (Florence, Italy)
  \emph{(\bibinfo{series}{CHI '08})}. \bibinfo{publisher}{Association for
  Computing Machinery}, \bibinfo{address}{New York, NY, USA},
  \bibinfo{pages}{1271–1274}.
\newblock
\showISBNx{9781605580111}
\href{https://doi.org/10.1145/1357054.1357252}{doi:\nolinkurl{10.1145/1357054.1357252}}


\bibitem[Ge et~al\mbox{.}(2026)]%
        {ge2026vocadesigningspeechcanvasinteraction}
\bibfield{author}{\bibinfo{person}{Yate Ge}, \bibinfo{person}{Run Yuan},
  \bibinfo{person}{Yueran Qi}, \bibinfo{person}{Wenjie He},
  \bibinfo{person}{Jiaqi Mo}, \bibinfo{person}{Yangshuo Chen},
  \bibinfo{person}{Wenbin Zuo}, \bibinfo{person}{Xiaohua Sun},
  \bibinfo{person}{Weiwei Guo}, {and} \bibinfo{person}{Qi Wang}.}
  \bibinfo{year}{2026}\natexlab{}.
\newblock \bibinfo{title}{VoCa: Designing Speech-Canvas Interaction for
  Voice-Based Conversational Agents}.
\newblock
\showeprint[arxiv]{2610.04706}~[cs.HC]
\urldef\tempurl%
\url{https://arxiv.org/abs/2610.04706}
\showURL{%
\tempurl}


\bibitem[Gonz{\'a}lez and Mark(2004)]%
        {gonzalez2004multitasking}
\bibfield{author}{\bibinfo{person}{Victor~M. Gonz{\'a}lez} {and}
  \bibinfo{person}{Gloria Mark}.} \bibinfo{year}{2004}\natexlab{}.
\newblock \showarticletitle{"Constant, constant, multi-tasking craziness":
  managing multiple working spheres}. In \bibinfo{booktitle}{\emph{Proceedings
  of the SIGCHI Conference on Human Factors in Computing Systems}} (Vienna,
  Austria) \emph{(\bibinfo{series}{CHI '04})}. \bibinfo{publisher}{Association
  for Computing Machinery}, \bibinfo{address}{New York, NY, USA},
  \bibinfo{pages}{113–120}.
\newblock
\showISBNx{1581137028}
\href{https://doi.org/10.1145/985692.985707}{doi:\nolinkurl{10.1145/985692.985707}}


\bibitem[Heer and Robertson(2007)]%
        {heer2007animatedtransitions}
\bibfield{author}{\bibinfo{person}{Jeffrey Heer} {and} \bibinfo{person}{George
  Robertson}.} \bibinfo{year}{2007}\natexlab{}.
\newblock \showarticletitle{Animated Transitions in Statistical Data Graphics}.
\newblock \bibinfo{journal}{\emph{IEEE Transactions on Visualization and
  Computer Graphics}} \bibinfo{volume}{13}, \bibinfo{number}{6}
  (\bibinfo{year}{2007}), \bibinfo{pages}{1240--1247}.
\newblock
\href{https://doi.org/10.1109/TVCG.2007.70539}{doi:\nolinkurl{10.1109/TVCG.2007.70539}}


\bibitem[Iqbal and Horvitz(2007)]%
        {iqbal2007disruption}
\bibfield{author}{\bibinfo{person}{Shamsi~T. Iqbal} {and} \bibinfo{person}{Eric
  Horvitz}.} \bibinfo{year}{2007}\natexlab{}.
\newblock \showarticletitle{Disruption and recovery of computing tasks: field
  study, analysis, and directions}. In \bibinfo{booktitle}{\emph{Proceedings of
  the SIGCHI Conference on Human Factors in Computing Systems}} (San Jose,
  California, USA) \emph{(\bibinfo{series}{CHI '07})}.
  \bibinfo{publisher}{Association for Computing Machinery},
  \bibinfo{address}{New York, NY, USA}, \bibinfo{pages}{677–686}.
\newblock
\showISBNx{9781595935939}
\href{https://doi.org/10.1145/1240624.1240730}{doi:\nolinkurl{10.1145/1240624.1240730}}


\bibitem[Janaka et~al\mbox{.}(2023)]%
        {janaka2023glassmessaging}
\bibfield{author}{\bibinfo{person}{Nuwan Janaka}, \bibinfo{person}{Jie Gao},
  \bibinfo{person}{Lin Zhu}, \bibinfo{person}{Shengdong Zhao},
  \bibinfo{person}{Lan Lyu}, \bibinfo{person}{Peisen Xu},
  \bibinfo{person}{Maximilian Nabokow}, \bibinfo{person}{Silang Wang}, {and}
  \bibinfo{person}{Yanch Ong}.} \bibinfo{year}{2023}\natexlab{}.
\newblock \showarticletitle{GlassMessaging: Towards Ubiquitous Messaging Using
  OHMDs}.
\newblock \bibinfo{journal}{\emph{Proc. ACM Interact. Mob. Wearable Ubiquitous
  Technol.}} \bibinfo{volume}{7}, \bibinfo{number}{3}, Article
  \bibinfo{articleno}{100} (\bibinfo{date}{Sept.} \bibinfo{year}{2023}),
  \bibinfo{numpages}{32}~pages.
\newblock
\href{https://doi.org/10.1145/3610931}{doi:\nolinkurl{10.1145/3610931}}


\bibitem[Jiang et~al\mbox{.}(2023)]%
        {jiang2023graphologue}
\bibfield{author}{\bibinfo{person}{Peiling Jiang}, \bibinfo{person}{Jude
  Rayan}, \bibinfo{person}{Steven~P. Dow}, {and} \bibinfo{person}{Haijun Xia}.}
  \bibinfo{year}{2023}\natexlab{}.
\newblock \showarticletitle{Graphologue: Exploring Large Language Model
  Responses with Interactive Diagrams}. In
  \bibinfo{booktitle}{\emph{Proceedings of the 36th Annual ACM Symposium on
  User Interface Software and Technology}} (San Francisco, CA, USA)
  \emph{(\bibinfo{series}{UIST '23})}. \bibinfo{publisher}{Association for
  Computing Machinery}, \bibinfo{address}{New York, NY, USA}, Article
  \bibinfo{articleno}{3}, \bibinfo{numpages}{20}~pages.
\newblock
\showISBNx{9798400701320}
\href{https://doi.org/10.1145/3586183.3606737}{doi:\nolinkurl{10.1145/3586183.3606737}}


\bibitem[Kim et~al\mbox{.}(2020)]%
        {kim2020interruptibilityinvehiclemultitasking}
\bibfield{author}{\bibinfo{person}{Auk Kim}, \bibinfo{person}{Jung-Mi Park},
  {and} \bibinfo{person}{Uichin Lee}.} \bibinfo{year}{2020}\natexlab{}.
\newblock \showarticletitle{Interruptibility for In-vehicle Multitasking:
  Influence of Voice Task Demands and Adaptive Behaviors}.
\newblock \bibinfo{journal}{\emph{Proc. ACM Interact. Mob. Wearable Ubiquitous
  Technol.}} \bibinfo{volume}{4}, \bibinfo{number}{1}, Article
  \bibinfo{articleno}{14} (\bibinfo{date}{March} \bibinfo{year}{2020}),
  \bibinfo{numpages}{22}~pages.
\newblock
\href{https://doi.org/10.1145/3381009}{doi:\nolinkurl{10.1145/3381009}}


\bibitem[Kim et~al\mbox{.}(2011)]%
        {kim2011usabilityproblemsinvoicesystems}
\bibfield{author}{\bibinfo{person}{Hee-Cheol Kim}, \bibinfo{person}{Deyun Liu},
  {and} \bibinfo{person}{Ho-Won Kim}.} \bibinfo{year}{2011}\natexlab{}.
\newblock \showarticletitle{Inherent Usability Problems in Interactive Voice
  Response Systems}. In \bibinfo{booktitle}{\emph{Human-Computer Interaction.
  Users and Applications}}, \bibfield{editor}{\bibinfo{person}{Julie~A. Jacko}}
  (Ed.). \bibinfo{publisher}{Springer Berlin Heidelberg},
  \bibinfo{address}{Berlin, Heidelberg}, \bibinfo{pages}{476--483}.
\newblock
\showISBNx{978-3-642-21619-0}


\bibitem[Kirsh and Maglio(1994)]%
        {kirsh1994epistemicpragmaticaction}
\bibfield{author}{\bibinfo{person}{David Kirsh} {and} \bibinfo{person}{Paul
  Maglio}.} \bibinfo{year}{1994}\natexlab{}.
\newblock \showarticletitle{On Distinguishing Epistemic from Pragmatic Action}.
\newblock \bibinfo{journal}{\emph{Cognitive Science}} \bibinfo{volume}{18},
  \bibinfo{number}{4} (\bibinfo{year}{1994}), \bibinfo{pages}{513--549}.
\newblock
\href{https://doi.org/10.1207/s15516709cog1804\_1}{doi:\nolinkurl{10.1207/s15516709cog1804\_1}}


\bibitem[Klokmose et~al\mbox{.}(2015)]%
        {klokmose2015webstrates}
\bibfield{author}{\bibinfo{person}{Clemens~N. Klokmose},
  \bibinfo{person}{James~R. Eagan}, \bibinfo{person}{Siemen Baader},
  \bibinfo{person}{Wendy Mackay}, {and} \bibinfo{person}{Michel
  Beaudouin-Lafon}.} \bibinfo{year}{2015}\natexlab{}.
\newblock \showarticletitle{Webstrates: Shareable Dynamic Media}. In
  \bibinfo{booktitle}{\emph{Proceedings of the 28th Annual ACM Symposium on
  User Interface Software \& Technology}} (Charlotte, NC, USA)
  \emph{(\bibinfo{series}{UIST '15})}. \bibinfo{publisher}{Association for
  Computing Machinery}, \bibinfo{address}{New York, NY, USA},
  \bibinfo{pages}{280–290}.
\newblock
\showISBNx{9781450337793}
\href{https://doi.org/10.1145/2807442.2807446}{doi:\nolinkurl{10.1145/2807442.2807446}}


\bibitem[Lakens(2022)]%
        {lakens2022samplesizejustification}
\bibfield{author}{\bibinfo{person}{Daniël Lakens}.}
  \bibinfo{year}{2022}\natexlab{}.
\newblock \showarticletitle{Sample Size Justification}.
\newblock \bibinfo{journal}{\emph{Collabra: Psychology}} \bibinfo{volume}{8},
  \bibinfo{number}{1} (\bibinfo{date}{03} \bibinfo{year}{2022}),
  \bibinfo{pages}{33267}.
\newblock
\showISSN{2474-7394}
\href{https://doi.org/10.1525/collabra.33267}{doi:\nolinkurl{10.1525/collabra.33267}}


\bibitem[Larsen-Ledet et~al\mbox{.}(2025)]%
        {larsenledet2025traces}
\bibfield{author}{\bibinfo{person}{Ida Larsen-Ledet}, \bibinfo{person}{Myriam
  Lewkowicz}, \bibinfo{person}{Clemens~N. Klokmose}, \bibinfo{person}{Carol
  Linehan}, {and} \bibinfo{person}{Luigina Ciolfi}.}
  \bibinfo{year}{2025}\natexlab{}.
\newblock \showarticletitle{Traces, Breadcrumbs, and Patina: Exploring and
  Designing with Traces of Activity}. In \bibinfo{booktitle}{\emph{Companion
  Publication of the 2025 Conference on Computer-Supported Cooperative Work and
  Social Computing}} (Bergen, Norway) \emph{(\bibinfo{series}{CSCW Companion
  '25})}. \bibinfo{publisher}{Association for Computing Machinery},
  \bibinfo{address}{New York, NY, USA}, \bibinfo{pages}{116–119}.
\newblock
\showISBNx{9798400714801}
\href{https://doi.org/10.1145/3715070.3748288}{doi:\nolinkurl{10.1145/3715070.3748288}}


\bibitem[Leahy and Sweller(2011)]%
        {leahy2011clt_transient}
\bibfield{author}{\bibinfo{person}{Wayne Leahy} {and} \bibinfo{person}{John
  Sweller}.} \bibinfo{year}{2011}\natexlab{}.
\newblock \showarticletitle{Cognitive load theory, modality of presentation and
  the transient information effect}.
\newblock \bibinfo{journal}{\emph{Applied Cognitive Psychology}}
  \bibinfo{volume}{25}, \bibinfo{number}{6} (\bibinfo{year}{2011}),
  \bibinfo{pages}{943--951}.
\newblock
\href{https://doi.org/10.1002/acp.1787}{doi:\nolinkurl{10.1002/acp.1787}}


\bibitem[Li et~al\mbox{.}(2026)]%
        {li2026orality}
\bibfield{author}{\bibinfo{person}{Wengxi Li}, \bibinfo{person}{Jingze Tian},
  {and} \bibinfo{person}{Can Liu}.} \bibinfo{year}{2026}\natexlab{}.
\newblock \showarticletitle{Orality: A Semantic Canvas for Externalizing and
  Clarifying Thoughts with Speech}. In \bibinfo{booktitle}{\emph{Proceedings of
  the 2026 CHI Conference on Human Factors in Computing Systems}} (Barcelona,
  Spain) \emph{(\bibinfo{series}{CHI '26})}. \bibinfo{publisher}{Association
  for Computing Machinery}, \bibinfo{address}{New York, NY, USA}, Article
  \bibinfo{articleno}{884}, \bibinfo{numpages}{29}~pages.
\newblock
\showISBNx{9798400722783}
\href{https://doi.org/10.1145/3772318.3791713}{doi:\nolinkurl{10.1145/3772318.3791713}}


\bibitem[Lin et~al\mbox{.}(2025)]%
        {lin2025fullduplexbench}
\bibfield{author}{\bibinfo{person}{Guan-Ting Lin}, \bibinfo{person}{Jiachen
  Lian}, \bibinfo{person}{Tingle Li}, \bibinfo{person}{Qirui Wang},
  \bibinfo{person}{Gopala Anumanchipalli}, \bibinfo{person}{Alexander~H. Liu},
  {and} \bibinfo{person}{Hung-Yi Lee}.} \bibinfo{year}{2025}\natexlab{}.
\newblock \showarticletitle{Full-Duplex-Bench: A Benchmark to Evaluate
  Full-Duplex Spoken Dialogue Models on Turn-taking Capabilities}. In
  \bibinfo{booktitle}{\emph{2025 IEEE Automatic Speech Recognition and
  Understanding Workshop (ASRU)}}. \bibinfo{publisher}{IEEE},
  \bibinfo{pages}{1--8}.
\newblock
\href{https://doi.org/10.1109/ASRU65441.2025.11433838}{doi:\nolinkurl{10.1109/ASRU65441.2025.11433838}}


\bibitem[Lin et~al\mbox{.}(2024)]%
        {lin2024rambler}
\bibfield{author}{\bibinfo{person}{Susan Lin}, \bibinfo{person}{Jeremy Warner},
  \bibinfo{person}{J.D. Zamfirescu-Pereira}, \bibinfo{person}{Matthew~G Lee},
  \bibinfo{person}{Sauhard Jain}, \bibinfo{person}{Shanqing Cai},
  \bibinfo{person}{Piyawat Lertvittayakumjorn}, \bibinfo{person}{Michael~Xuelin
  Huang}, \bibinfo{person}{Shumin Zhai}, \bibinfo{person}{Bjoern Hartmann},
  {and} \bibinfo{person}{Can Liu}.} \bibinfo{year}{2024}\natexlab{}.
\newblock \showarticletitle{Rambler: Supporting Writing With Speech via
  LLM-Assisted Gist Manipulation}. In \bibinfo{booktitle}{\emph{Proceedings of
  the 2024 CHI Conference on Human Factors in Computing Systems}} (Honolulu,
  HI, USA) \emph{(\bibinfo{series}{CHI '24})}. \bibinfo{publisher}{Association
  for Computing Machinery}, \bibinfo{address}{New York, NY, USA}, Article
  \bibinfo{articleno}{1043}, \bibinfo{numpages}{19}~pages.
\newblock
\showISBNx{9798400703300}
\href{https://doi.org/10.1145/3613904.3642217}{doi:\nolinkurl{10.1145/3613904.3642217}}


\bibitem[Liu et~al\mbox{.}(2023)]%
        {liu2023captionstability}
\bibfield{author}{\bibinfo{person}{Xingyu~``Bruce'' Liu}, \bibinfo{person}{Jun
  Zhang}, \bibinfo{person}{Leonardo Ferrer}, \bibinfo{person}{Susan Xu},
  \bibinfo{person}{Vikas Bahirwani}, \bibinfo{person}{Boris Smus},
  \bibinfo{person}{Alex Olwal}, {and} \bibinfo{person}{Ruofei Du}.}
  \bibinfo{year}{2023}\natexlab{}.
\newblock \showarticletitle{Modeling and Improving Text Stability in Live
  Captions}. In \bibinfo{booktitle}{\emph{Extended Abstracts of the 2023 CHI
  Conference on Human Factors in Computing Systems}}
  \emph{(\bibinfo{series}{CHI EA '23})}. \bibinfo{publisher}{ACM},
  \bibinfo{address}{New York, NY, USA}, \bibinfo{pages}{1--9}.
\newblock
\href{https://doi.org/10.1145/3544549.3585609}{doi:\nolinkurl{10.1145/3544549.3585609}}


\bibitem[MacLeod et~al\mbox{.}(2010)]%
        {macleod2010theproductioneffect}
\bibfield{author}{\bibinfo{person}{Colin MacLeod}, \bibinfo{person}{Nigel
  Gopie}, \bibinfo{person}{Kathleen Hourihan}, \bibinfo{person}{Karen Neary},
  {and} \bibinfo{person}{Jason Ozubko}.} \bibinfo{year}{2010}\natexlab{}.
\newblock \showarticletitle{The Production Effect: Delineation of a
  Phenomenon}.
\newblock \bibinfo{journal}{\emph{Journal of Experimental Psychology: Learning,
  Memory, and Cognition}}  \bibinfo{volume}{36} (\bibinfo{date}{05}
  \bibinfo{year}{2010}), \bibinfo{pages}{671--685}.
\newblock
\href{https://doi.org/10.1037/a0018785}{doi:\nolinkurl{10.1037/a0018785}}


\bibitem[Mark et~al\mbox{.}(2008)]%
        {mark2008interruptedwork}
\bibfield{author}{\bibinfo{person}{Gloria Mark}, \bibinfo{person}{Daniela
  Gudith}, {and} \bibinfo{person}{Ulrich Klocke}.}
  \bibinfo{year}{2008}\natexlab{}.
\newblock \showarticletitle{The cost of interrupted work: more speed and
  stress}. In \bibinfo{booktitle}{\emph{Proceedings of the SIGCHI Conference on
  Human Factors in Computing Systems}} (Florence, Italy)
  \emph{(\bibinfo{series}{CHI '08})}. \bibinfo{publisher}{Association for
  Computing Machinery}, \bibinfo{address}{New York, NY, USA},
  \bibinfo{pages}{107–110}.
\newblock
\showISBNx{9781605580111}
\href{https://doi.org/10.1145/1357054.1357072}{doi:\nolinkurl{10.1145/1357054.1357072}}


\bibitem[Matthews(2006)]%
        {matthews2006glanceable}
\bibfield{author}{\bibinfo{person}{Tara Matthews}.}
  \bibinfo{year}{2006}\natexlab{}.
\newblock \showarticletitle{Designing and evaluating glanceable peripheral
  displays}. In \bibinfo{booktitle}{\emph{Proceedings of the 6th Conference on
  Designing Interactive Systems}} (University Park, PA, USA)
  \emph{(\bibinfo{series}{DIS '06})}. \bibinfo{publisher}{Association for
  Computing Machinery}, \bibinfo{address}{New York, NY, USA},
  \bibinfo{pages}{343–345}.
\newblock
\showISBNx{1595933670}
\href{https://doi.org/10.1145/1142405.1142457}{doi:\nolinkurl{10.1145/1142405.1142457}}


\bibitem[Monk et~al\mbox{.}(2026)]%
        {monk2026incarllm}
\bibfield{author}{\bibinfo{person}{Chris Monk}, \bibinfo{person}{Allegra
  Ayala}, \bibinfo{person}{Christine S.~P. Yu}, \bibinfo{person}{Gregory~M.
  Fitch}, {and} \bibinfo{person}{Dara Gruber}.}
  \bibinfo{year}{2026}\natexlab{}.
\newblock \bibinfo{title}{Visual and Cognitive Demands of a Large Language
  Model-Powered In-vehicle Conversational Agent}.
\newblock
\showeprint[arxiv]{2601.15034}
\urldef\tempurl%
\url{https://arxiv.org/abs/2601.15034}
\showURL{%
\tempurl}


\bibitem[Monk et~al\mbox{.}(2008)]%
        {monk2008interruptionduration}
\bibfield{author}{\bibinfo{person}{Christopher~A. Monk},
  \bibinfo{person}{J.~Gregory Trafton}, {and} \bibinfo{person}{Deborah~A.
  Boehm-Davis}.} \bibinfo{year}{2008}\natexlab{}.
\newblock \showarticletitle{The effect of interruption duration and demand on
  resuming suspended goals.}
\newblock \bibinfo{journal}{\emph{Journal of Experimental Psychology: Applied}}
  \bibinfo{volume}{14}, \bibinfo{number}{4} (\bibinfo{year}{2008}),
  \bibinfo{pages}{299--313}.
\newblock
\showISSN{1076-898X}
\href{https://doi.org/10.1037/a0014402}{doi:\nolinkurl{10.1037/a0014402}}


\bibitem[Moore et~al\mbox{.}(2024)]%
        {moore2024understandingtwowaystreet}
\bibfield{author}{\bibinfo{person}{Robert~J. Moore}, \bibinfo{person}{Sungeun
  An}, {and} \bibinfo{person}{Olivia~H. Marrese}.}
  \bibinfo{year}{2024}\natexlab{}.
\newblock \showarticletitle{Understanding is a Two-Way Street: User-Initiated
  Repair on Agent Responses and Hearing in Conversational Interfaces}.
\newblock \bibinfo{journal}{\emph{Proc. ACM Hum.-Comput. Interact.}}
  \bibinfo{volume}{8}, \bibinfo{number}{CSCW1}, Article
  \bibinfo{articleno}{187} (\bibinfo{date}{April} \bibinfo{year}{2024}),
  \bibinfo{numpages}{26}~pages.
\newblock
\href{https://doi.org/10.1145/3641026}{doi:\nolinkurl{10.1145/3641026}}


\bibitem[Neustaedter and Sengers(2012)]%
        {neustaedter2012autobiographicaldesign}
\bibfield{author}{\bibinfo{person}{Carman Neustaedter} {and}
  \bibinfo{person}{Phoebe Sengers}.} \bibinfo{year}{2012}\natexlab{}.
\newblock \showarticletitle{Autobiographical design in HCI research: designing
  and learning through use-it-yourself}. In
  \bibinfo{booktitle}{\emph{Proceedings of the Designing Interactive Systems
  Conference}} (Newcastle Upon Tyne, United Kingdom)
  \emph{(\bibinfo{series}{DIS '12})}. \bibinfo{publisher}{Association for
  Computing Machinery}, \bibinfo{address}{New York, NY, USA},
  \bibinfo{pages}{514–523}.
\newblock
\showISBNx{9781450312103}
\href{https://doi.org/10.1145/2317956.2318034}{doi:\nolinkurl{10.1145/2317956.2318034}}


\bibitem[Olsen(2007)]%
        {olsen2007evaluation}
\bibfield{author}{\bibinfo{person}{Dan~R. Olsen, Jr.}}
  \bibinfo{year}{2007}\natexlab{}.
\newblock \showarticletitle{Evaluating user interface systems research}. In
  \bibinfo{booktitle}{\emph{Proceedings of the 20th Annual ACM Symposium on
  User Interface Software \& Technology}} (Newport, RI, USA)
  \emph{(\bibinfo{series}{UIST '07})}. \bibinfo{publisher}{Association for
  Computing Machinery}, \bibinfo{address}{New York, NY, USA},
  \bibinfo{pages}{251--258}.
\newblock
\href{https://doi.org/10.1145/1294211.1294256}{doi:\nolinkurl{10.1145/1294211.1294256}}


\bibitem[Park et~al\mbox{.}(2025)]%
        {park2025visualbackchannels}
\bibfield{author}{\bibinfo{person}{Yoonvin Park}, \bibinfo{person}{Doha Kim},
  {and} \bibinfo{person}{Hayeon Song}.} \bibinfo{year}{2025}\natexlab{}.
\newblock \showarticletitle{I Know You're Listening: Designing Visual
  Backchannels for Voice User Interfaces}. In
  \bibinfo{booktitle}{\emph{Companion Proceedings of the 30th International
  Conference on Intelligent User Interfaces}} (Cagliari, Italy)
  \emph{(\bibinfo{series}{IUI '25 Companion})}. \bibinfo{publisher}{Association
  for Computing Machinery}, \bibinfo{address}{New York, NY, USA},
  \bibinfo{pages}{56–59}.
\newblock
\showISBNx{9798400714092}
\href{https://doi.org/10.1145/3708557.3716343}{doi:\nolinkurl{10.1145/3708557.3716343}}


\bibitem[Ram and Zhao(2021)]%
        {ram2021onthegovideolearning}
\bibfield{author}{\bibinfo{person}{Ashwin Ram} {and} \bibinfo{person}{Shengdong
  Zhao}.} \bibinfo{year}{2021}\natexlab{}.
\newblock \showarticletitle{LSVP: Towards Effective On-the-go Video Learning
  Using Optical Head-Mounted Displays}.
\newblock \bibinfo{journal}{\emph{Proc. ACM Interact. Mob. Wearable Ubiquitous
  Technol.}} \bibinfo{volume}{5}, \bibinfo{number}{1}, Article
  \bibinfo{articleno}{30} (\bibinfo{date}{March} \bibinfo{year}{2021}),
  \bibinfo{numpages}{27}~pages.
\newblock
\href{https://doi.org/10.1145/3448118}{doi:\nolinkurl{10.1145/3448118}}


\bibitem[Shi et~al\mbox{.}(2026)]%
        {shi2026talksketch}
\bibfield{author}{\bibinfo{person}{Weiyan Shi}, \bibinfo{person}{Sunaya
  Upadhyay}, \bibinfo{person}{Geraldine Quek}, {and} \bibinfo{person}{Kenny
  Tsu~Wei Choo}.} \bibinfo{year}{2026}\natexlab{}.
\newblock \showarticletitle{TalkSketch: Multimodal Generative AI for Real-Time
  Sketch Ideation with Speech}. In \bibinfo{booktitle}{\emph{Creative AI for
  Live Interactive Performances}}, \bibfield{editor}{\bibinfo{person}{Kieran
  Woodward}, \bibinfo{person}{Alicia Falcon-Caro}, \bibinfo{person}{Richard
  Ramchurn}, {and} \bibinfo{person}{Steve Benford}} (Eds.).
  \bibinfo{publisher}{Springer Nature Switzerland}, \bibinfo{address}{Cham},
  \bibinfo{pages}{83--97}.
\newblock


\bibitem[Shneiderman(1996)]%
        {shneiderman1996eyeshaveit}
\bibfield{author}{\bibinfo{person}{B. Shneiderman}.}
  \bibinfo{year}{1996}\natexlab{}.
\newblock \showarticletitle{The eyes have it: a task by data type taxonomy for
  information visualizations}. In \bibinfo{booktitle}{\emph{Proceedings 1996
  IEEE Symposium on Visual Languages}}. \bibinfo{publisher}{IEEE},
  \bibinfo{pages}{336--343}.
\newblock
\href{https://doi.org/10.1109/VL.1996.545307}{doi:\nolinkurl{10.1109/VL.1996.545307}}


\bibitem[Simons and Rensink(2005)]%
        {simons2005changeblindness}
\bibfield{author}{\bibinfo{person}{Daniel~J. Simons} {and}
  \bibinfo{person}{Ronald~A. Rensink}.} \bibinfo{year}{2005}\natexlab{}.
\newblock \showarticletitle{Change Blindness: Past, Present, and Future}.
\newblock \bibinfo{journal}{\emph{Trends in Cognitive Sciences}}
  \bibinfo{volume}{9}, \bibinfo{number}{1} (\bibinfo{year}{2005}),
  \bibinfo{pages}{16--20}.
\newblock
\href{https://doi.org/10.1016/j.tics.2004.11.006}{doi:\nolinkurl{10.1016/j.tics.2004.11.006}}


\bibitem[Skantze(2021)]%
        {skantze2021turntaking}
\bibfield{author}{\bibinfo{person}{Gabriel Skantze}.}
  \bibinfo{year}{2021}\natexlab{}.
\newblock \showarticletitle{Turn-taking in Conversational Systems and
  Human-Robot Interaction: A Review}.
\newblock \bibinfo{journal}{\emph{Computer Speech \& Language}}
  \bibinfo{volume}{67} (\bibinfo{year}{2021}), \bibinfo{pages}{101178}.
\newblock
\showISSN{0885-2308}
\href{https://doi.org/10.1016/j.csl.2020.101178}{doi:\nolinkurl{10.1016/j.csl.2020.101178}}


\bibitem[Slamecka and Graf(1978)]%
        {slamecka1978thegenerationeffect}
\bibfield{author}{\bibinfo{person}{Norman~J. Slamecka} {and}
  \bibinfo{person}{Peter Graf}.} \bibinfo{year}{1978}\natexlab{}.
\newblock \showarticletitle{The generation effect: Delineation of a
  phenomenon.}
\newblock \bibinfo{journal}{\emph{Journal of Experimental Psychology: Human
  Learning and Memory}} \bibinfo{volume}{4}, \bibinfo{number}{6}
  (\bibinfo{date}{Nov.} \bibinfo{year}{1978}), \bibinfo{pages}{592--604}.
\newblock
\href{https://doi.org/10.1037/0278-7393.4.6.592}{doi:\nolinkurl{10.1037/0278-7393.4.6.592}}


\bibitem[Srinivasan et~al\mbox{.}(2025)]%
        {srinivasan2025attentionaware}
\bibfield{author}{\bibinfo{person}{Arvind Srinivasan},
  \bibinfo{person}{Johannes Ellemose}, \bibinfo{person}{Peter W.~S. Butcher},
  \bibinfo{person}{Panagiotis~D. Ritsos}, {and} \bibinfo{person}{Niklas
  Elmqvist}.} \bibinfo{year}{2025}\natexlab{}.
\newblock \showarticletitle{Attention-Aware Visualization: Tracking and
  Responding to User Perception Over Time}.
\newblock \bibinfo{journal}{\emph{IEEE Transactions on Visualization and
  Computer Graphics}} \bibinfo{volume}{31}, \bibinfo{number}{1}
  (\bibinfo{date}{Jan.} \bibinfo{year}{2025}), \bibinfo{pages}{1017–1027}.
\newblock
\showISSN{1077-2626}
\href{https://doi.org/10.1109/TVCG.2024.3456300}{doi:\nolinkurl{10.1109/TVCG.2024.3456300}}


\bibitem[Stark et~al\mbox{.}(2005)]%
        {stark2005appropriation}
\bibfield{author}{\bibinfo{person}{Louisa-Jayne Stark},
  \bibinfo{person}{Timothy Perfect}, {and} \bibinfo{person}{Stephen Newstead}.}
  \bibinfo{year}{2005}\natexlab{}.
\newblock \showarticletitle{When elaboration leads to appropriation:
  Unconscious plagiarism in a creative task}.
\newblock \bibinfo{journal}{\emph{Memory}} \bibinfo{volume}{13},
  \bibinfo{number}{6} (\bibinfo{year}{2005}), \bibinfo{pages}{561--573}.
\newblock
\href{https://doi.org/10.1080/09658210444000232}{doi:\nolinkurl{10.1080/09658210444000232}}


\bibitem[Stifelman et~al\mbox{.}(1993)]%
        {stifelman1993voicenotes}
\bibfield{author}{\bibinfo{person}{Lisa~J. Stifelman}, \bibinfo{person}{Barry
  Arons}, \bibinfo{person}{Chris Schmandt}, {and} \bibinfo{person}{Eric~A.
  Hulteen}.} \bibinfo{year}{1993}\natexlab{}.
\newblock \showarticletitle{VoiceNotes: a speech interface for a hand-held
  voice notetaker}. In \bibinfo{booktitle}{\emph{Proceedings of the INTERACT
  '93 and CHI '93 Conference on Human Factors in Computing Systems}}
  (Amsterdam, The Netherlands) \emph{(\bibinfo{series}{CHI '93})}.
  \bibinfo{publisher}{Association for Computing Machinery},
  \bibinfo{address}{New York, NY, USA}, \bibinfo{pages}{179–186}.
\newblock
\showISBNx{0897915755}
\href{https://doi.org/10.1145/169059.169150}{doi:\nolinkurl{10.1145/169059.169150}}


\bibitem[Strayer and Johnston(2001)]%
        {strayer2001distraction}
\bibfield{author}{\bibinfo{person}{David~L. Strayer} {and}
  \bibinfo{person}{William~A. Johnston}.} \bibinfo{year}{2001}\natexlab{}.
\newblock \showarticletitle{Driven to Distraction: Dual-Task Studies of
  Simulated Driving and Conversing on a Cellular Telephone}.
\newblock \bibinfo{journal}{\emph{Psychological Science}} \bibinfo{volume}{12},
  \bibinfo{number}{6} (\bibinfo{date}{Nov.} \bibinfo{year}{2001}),
  \bibinfo{pages}{462--466}.
\newblock
\showISSN{1467-9280}
\href{https://doi.org/10.1111/1467-9280.00386}{doi:\nolinkurl{10.1111/1467-9280.00386}}


\bibitem[Suh et~al\mbox{.}(2023)]%
        {suh2023sensecape}
\bibfield{author}{\bibinfo{person}{Sangho Suh}, \bibinfo{person}{Bryan Min},
  \bibinfo{person}{Srishti Palani}, {and} \bibinfo{person}{Haijun Xia}.}
  \bibinfo{year}{2023}\natexlab{}.
\newblock \showarticletitle{Sensecape: Enabling Multilevel Exploration and
  Sensemaking with Large Language Models}. In
  \bibinfo{booktitle}{\emph{Proceedings of the 36th Annual ACM Symposium on
  User Interface Software and Technology}} (San Francisco, CA, USA)
  \emph{(\bibinfo{series}{UIST '23})}. \bibinfo{publisher}{Association for
  Computing Machinery}, \bibinfo{address}{New York, NY, USA}, Article
  \bibinfo{articleno}{1}, \bibinfo{numpages}{18}~pages.
\newblock
\showISBNx{9798400701320}
\href{https://doi.org/10.1145/3586183.3606756}{doi:\nolinkurl{10.1145/3586183.3606756}}


\bibitem[Trafton et~al\mbox{.}(2003)]%
        {trafton2003preparingtoresume}
\bibfield{author}{\bibinfo{person}{J.~Gregory Trafton},
  \bibinfo{person}{Erik~M. Altmann}, \bibinfo{person}{Derek~P. Brock}, {and}
  \bibinfo{person}{Farilee~E. Mintz}.} \bibinfo{year}{2003}\natexlab{}.
\newblock \showarticletitle{Preparing to resume an interrupted task: effects of
  prospective goal encoding and retrospective rehearsal}.
\newblock \bibinfo{journal}{\emph{Int. J. Hum.-Comput. Stud.}}
  \bibinfo{volume}{58}, \bibinfo{number}{5} (\bibinfo{date}{May}
  \bibinfo{year}{2003}), \bibinfo{pages}{583–603}.
\newblock
\showISSN{1071-5819}
\href{https://doi.org/10.1016/S1071-5819(03)00023-5}{doi:\nolinkurl{10.1016/S1071-5819(03)00023-5}}


\bibitem[Tsai et~al\mbox{.}(2025)]%
        {tsai2025gazenoter}
\bibfield{author}{\bibinfo{person}{Hsin-Ruey Tsai}, \bibinfo{person}{Shih-Kang
  Chiu}, {and} \bibinfo{person}{Bryan Wang}.} \bibinfo{year}{2025}\natexlab{}.
\newblock \showarticletitle{GazeNoter: Co-Piloted AR Note-Taking via Gaze
  Selection of LLM Suggestions to Match Users' Intentions}. In
  \bibinfo{booktitle}{\emph{Proceedings of the 2025 CHI Conference on Human
  Factors in Computing Systems}} \emph{(\bibinfo{series}{CHI '25})}.
  \bibinfo{publisher}{Association for Computing Machinery},
  \bibinfo{address}{New York, NY, USA}, Article \bibinfo{articleno}{32},
  \bibinfo{numpages}{22}~pages.
\newblock
\showISBNx{9798400713941}
\href{https://doi.org/10.1145/3706598.3714294}{doi:\nolinkurl{10.1145/3706598.3714294}}


\bibitem[Vi{\'e}gas and Donath(1999)]%
        {viegas1999chatcircles}
\bibfield{author}{\bibinfo{person}{Fernanda~B. Vi{\'e}gas} {and}
  \bibinfo{person}{Judith~S. Donath}.} \bibinfo{year}{1999}\natexlab{}.
\newblock \showarticletitle{Chat circles}. In
  \bibinfo{booktitle}{\emph{Proceedings of the SIGCHI Conference on Human
  Factors in Computing Systems}} (Pittsburgh, Pennsylvania, USA)
  \emph{(\bibinfo{series}{CHI '99})}. \bibinfo{publisher}{Association for
  Computing Machinery}, \bibinfo{address}{New York, NY, USA},
  \bibinfo{pages}{9–16}.
\newblock
\showISBNx{0201485591}
\href{https://doi.org/10.1145/302979.302981}{doi:\nolinkurl{10.1145/302979.302981}}


\bibitem[Wickens(2008)]%
        {wickens2008multipleresources}
\bibfield{author}{\bibinfo{person}{Christopher~D. Wickens}.}
  \bibinfo{year}{2008}\natexlab{}.
\newblock \showarticletitle{Multiple Resources and Mental Workload}.
\newblock \bibinfo{journal}{\emph{Human Factors: The Journal of the Human
  Factors and Ergonomics Society}} \bibinfo{volume}{50}, \bibinfo{number}{3}
  (\bibinfo{date}{June} \bibinfo{year}{2008}), \bibinfo{pages}{449--455}.
\newblock
\showISSN{1547-8181}
\href{https://doi.org/10.1518/001872008x288394}{doi:\nolinkurl{10.1518/001872008x288394}}


\bibitem[Wu and Yao(2026)]%
        {wu2026humanagencyintheageofconversationalai}
\bibfield{author}{\bibinfo{person}{Mengke Wu} {and} \bibinfo{person}{Mike
  Yao}.} \bibinfo{year}{2026}\natexlab{}.
\newblock \showarticletitle{After the Interface: Relocating Human Agency in the
  Age of Conversational AI}. In \bibinfo{booktitle}{\emph{Proceedings of the
  8th ACM Conference on Conversational User Interfaces}} (Bremen, Germany)
  \emph{(\bibinfo{series}{CUI '26})}. \bibinfo{publisher}{Association for
  Computing Machinery}, \bibinfo{address}{New York, NY, USA}, Article
  \bibinfo{articleno}{49}, \bibinfo{numpages}{7}~pages.
\newblock
\showISBNx{9798400727412}
\href{https://doi.org/10.1145/3816046.3816301}{doi:\nolinkurl{10.1145/3816046.3816301}}


\bibitem[Yang et~al\mbox{.}(2024)]%
        {yang2024talk2care}
\bibfield{author}{\bibinfo{person}{Ziqi Yang}, \bibinfo{person}{Xuhai Xu},
  \bibinfo{person}{Bingsheng Yao}, \bibinfo{person}{Ethan Rogers},
  \bibinfo{person}{Shao Zhang}, \bibinfo{person}{Stephen Intille},
  \bibinfo{person}{Nawar Shara}, \bibinfo{person}{Guodong~Gordon Gao}, {and}
  \bibinfo{person}{Dakuo Wang}.} \bibinfo{year}{2024}\natexlab{}.
\newblock \showarticletitle{Talk2Care: An LLM-based Voice Assistant for
  Communication between Healthcare Providers and Older Adults}.
\newblock \bibinfo{journal}{\emph{Proc. ACM Interact. Mob. Wearable Ubiquitous
  Technol.}} \bibinfo{volume}{8}, \bibinfo{number}{2}, Article
  \bibinfo{articleno}{73} (\bibinfo{date}{May} \bibinfo{year}{2024}),
  \bibinfo{numpages}{35}~pages.
\newblock
\href{https://doi.org/10.1145/3659625}{doi:\nolinkurl{10.1145/3659625}}


\bibitem[Yen and Zhao(2024)]%
        {yen2024memolet}
\bibfield{author}{\bibinfo{person}{Ryan Yen} {and} \bibinfo{person}{Jian
  Zhao}.} \bibinfo{year}{2024}\natexlab{}.
\newblock \showarticletitle{Memolet: Reifying the Reuse of User-AI
  Conversational Memories}. In \bibinfo{booktitle}{\emph{Proceedings of the
  37th Annual ACM Symposium on User Interface Software and Technology}}
  (Pittsburgh, PA, USA) \emph{(\bibinfo{series}{UIST '24})}.
  \bibinfo{publisher}{Association for Computing Machinery},
  \bibinfo{address}{New York, NY, USA}, Article \bibinfo{articleno}{58},
  \bibinfo{numpages}{22}~pages.
\newblock
\showISBNx{9798400706288}
\href{https://doi.org/10.1145/3654777.3676388}{doi:\nolinkurl{10.1145/3654777.3676388}}


\bibitem[Zimmerman et~al\mbox{.}(2007)]%
        {zimmermann2007researchthroughdesign}
\bibfield{author}{\bibinfo{person}{John Zimmerman}, \bibinfo{person}{Jodi
  Forlizzi}, {and} \bibinfo{person}{Shelley Evenson}.}
  \bibinfo{year}{2007}\natexlab{}.
\newblock \showarticletitle{Research through design as a method for interaction
  design research in HCI}. In \bibinfo{booktitle}{\emph{Proceedings of the
  SIGCHI Conference on Human Factors in Computing Systems}} (San Jose,
  California, USA) \emph{(\bibinfo{series}{CHI '07})}.
  \bibinfo{publisher}{Association for Computing Machinery},
  \bibinfo{address}{New York, NY, USA}, \bibinfo{pages}{493–502}.
\newblock
\showISBNx{9781595935939}
\href{https://doi.org/10.1145/1240624.1240704}{doi:\nolinkurl{10.1145/1240624.1240704}}


\bibitem[Zindulka et~al\mbox{.}(2026)]%
        {zindulka2026aimemorygap}
\bibfield{author}{\bibinfo{person}{Tim Zindulka}, \bibinfo{person}{Sven
  Goller}, \bibinfo{person}{Daniela Fernandes}, \bibinfo{person}{Robin Welsch},
  {and} \bibinfo{person}{Daniel Buschek}.} \bibinfo{year}{2026}\natexlab{}.
\newblock \showarticletitle{The AI Memory Gap: Users Misremember What They
  Created With AI or Without}. In \bibinfo{booktitle}{\emph{Proceedings of the
  2026 CHI Conference on Human Factors in Computing Systems}} (Barcelona,
  Spain) \emph{(\bibinfo{series}{CHI '26})}. \bibinfo{publisher}{Association
  for Computing Machinery}, \bibinfo{address}{New York, NY, USA}, Article
  \bibinfo{articleno}{61}, \bibinfo{numpages}{22}~pages.
\newblock
\showISBNx{9798400722783}
\href{https://doi.org/10.1145/3772318.3791494}{doi:\nolinkurl{10.1145/3772318.3791494}}


\bibitem[Zulfikar et~al\mbox{.}(2024)]%
        {zulfikar2024memoro}
\bibfield{author}{\bibinfo{person}{Wazeer~Deen Zulfikar},
  \bibinfo{person}{Samantha Chan}, {and} \bibinfo{person}{Pattie Maes}.}
  \bibinfo{year}{2024}\natexlab{}.
\newblock \showarticletitle{Memoro: Using Large Language Models to Realize a
  Concise Interface for Real-Time Memory Augmentation}. In
  \bibinfo{booktitle}{\emph{Proceedings of the 2024 CHI Conference on Human
  Factors in Computing Systems}} (Honolulu, HI, USA)
  \emph{(\bibinfo{series}{CHI '24})}. \bibinfo{publisher}{Association for
  Computing Machinery}, \bibinfo{address}{New York, NY, USA}, Article
  \bibinfo{articleno}{450}, \bibinfo{numpages}{18}~pages.
\newblock
\showISBNx{9798400703300}
\href{https://doi.org/10.1145/3613904.3642450}{doi:\nolinkurl{10.1145/3613904.3642450}}


\end{thebibliography}

\appendix

\end{document}